\documentclass[lettersize,journal]{IEEEtran}

\usepackage{url}
\usepackage{color}

\usepackage{amsmath,amsfonts,amssymb}
\usepackage{bm}
\usepackage{algorithm}
\usepackage{algpseudocode}

\usepackage{wasysym}

\usepackage{multirow}
\usepackage{blindtext}
\usepackage{wrapfig}
\usepackage{booktabs}
\usepackage{subfigure}

\usepackage{tcolorbox}
\tcbuselibrary{breakable}
\usepackage{xcolor}
\usepackage{soul}

\usepackage{amsmath,amsfonts}
\usepackage{textcomp}
\usepackage{stfloats}
\usepackage{url}
\usepackage{verbatim}
\usepackage{graphicx}

\usepackage{textcomp}
\usepackage{graphicx}
\usepackage[shortlabels]{enumitem}
\usepackage{balance}
\usepackage{comment}
\usepackage{CJKutf8} 

\usepackage{url}
\usepackage{color}

\usepackage{bm}
\usepackage{algorithm}
\usepackage{algpseudocode}

\usepackage{wasysym}

\usepackage{multirow}

\usepackage{wrapfig}
\usepackage{booktabs}
\usepackage{subfigure}

\usepackage{tcolorbox}
\tcbuselibrary{breakable}
\usepackage{xcolor}
\usepackage{soul}  
\usepackage{balance}

\usepackage{colortbl}

\definecolor{lightpink}{rgb}{1.0, 0.87, 0.87}
\definecolor{lightpurple}{rgb}{0.94, 0.85, 0.94}
\definecolor{lightgreen}{rgb}{0.88, 1.0, 0.88}
\definecolor{lightyellow}{rgb}{1.0, 1.0, 0.88}
\definecolor{lightblue}{rgb}{0.88, 0.94, 1.0}
\definecolor{lightgrey}{rgb}{0.8, 0.8, 0.8}

\usepackage{listings}

\definecolor{codegreen}{rgb}{0,0.6,0}
\definecolor{codegray}{rgb}{0.5,0.5,0.5}
\definecolor{codepurple}{rgb}{0.58,0,0.82}
\definecolor{backcolour}{rgb}{0.95,0.95,0.92}

\lstdefinestyle{mystyle}{
    backgroundcolor=\color{backcolour},   
    commentstyle=\color{codegreen},
    keywordstyle=\color{magenta},
    numberstyle=\tiny\color{codegray},
    stringstyle=\color{codepurple},
    basicstyle=\footnotesize\ttfamily,
    breakatwhitespace=false,         
    breaklines=true,                 
    captionpos=b,                    
    keepspaces=true,                 
    numbers=left,                    
    numbersep=5pt,                  
    showspaces=false,                
    showstringspaces=false,
    showtabs=false,   
    tabsize=3,
    frame=tb,
    aboveskip=2mm,
    belowskip=2mm
}

\usepackage{hyperref}
\usepackage[hyphenbreaks]{breakurl}

\begin{document}

\title{\textsc{LSem2Vec}: An Effective Embedding Approach for Code Clone Detection and Clustering}

\author{Zixiang Xian, Rubing Huang, Chenhui Cui, Chunrong Fang, Zhenyu Chen
        \thanks{Zixiang Xian and Chenhui Cui are with the School of Computer Science and Engineering, Macau University of Science and Technology, Macao SAR 999078, China. E-mail: 3220001352@student.must.edu.mo, 3230002105@student.must.edu.mo.}
        \thanks{Rubing Huang is with the School of Computer Science and Engineering, Macau University of Science and Technology, Macao SAR 999078, China; and also with Macau University of Science and Technology Zhuhai MUST Science and Technology Research Institute, Zhuhai, Guangdong 519099, China. E-mail: rbhuang@must.edu.mo.}
        \thanks{Chunrong Fang and Zhenyu Chen are with the State Key Laboratory for Novel Software Technology, Nanjing University, Nanjing 210093. China, E-mail: fangchunrong@nju.edu.cn, zychen@nju.edu.cn.}
}

\maketitle

\begin{abstract}
The advent of large language models (LLMs) has significantly advanced artificial intelligence (AI) in software engineering (SE), with source code embeddings playing a crucial role in tasks such as source code clone detection and source code clustering. However, existing methods for source code embedding, including those based on LLMs, often rely on costly supervised training or fine-tuning for domain adaptation. This paper introduces CodeSemVec, a novel approach to embedding source code by combining large language models and sentence embedding models. CodeSemVec eliminates the need for task-specific training or fine-tuning and effectively addresses erroneous information commonly found in LLM-generated outputs. To evaluate the performance of CodeSemVec, we conducted a series of experiments across three datasets spanning different programming languages, using various LLMs and sentence embedding models. The experimental results demonstrate the effectiveness and superiority of CodeSemVec over state-of-the-art unsupervised methods, including SourcererCC, Code2vec, InferCode, TransformCode, and LLM2Vec. Our findings highlight the potential of CodeSemVec to advance the field of SE by providing robust and efficient solutions for source code embedding tasks.
\end{abstract}


\section{Introduction}
\label{sec_intro}

Over the past decade, artificial intelligence (AI) has gained significant traction in software engineering (SE). The integration of AI into SE practices aims to enhance the efficiency and effectiveness of software development processes, thereby boosting productivity and fostering innovation. A key milestone in this integration has been the advent of pre-trained models (PTMs). PTMs such as CodeBERT \cite{codebert2020} have been trained on large-scale source code datasets, demonstrating a deep understanding of various programming languages. 
Despite their impressive capabilities, PTMs often require fine-tuning on task-specific data to achieve optimal performance in specialized SE tasks. This fine-tuning process involves adapting the model to the specific requirements of the target task, ensuring it produces accurate and relevant results. While fine-tuning significantly enhances the model's performance, it demands substantial computational resources. Additionally, the need for large amounts of labeled data for fine-tuning poses a significant challenge to the use of PTMs.

The introduction of LLMs like GPT-3 \cite{brown2020language} and GPT-4 \cite{achiam2023gpt} from OpenAI, as well as GLM3 and GLM4 from Tsinghua University and Zhipu AI \cite{glm2024chatglm}, has greatly advanced the field. These LLMs are known for their impressive zero-shot learning abilities due to their vast pre-trained knowledge. This allows them to handle certain tasks without additional training or fine-tuning \cite{alshahwan2024assured}, a significant advantage over PTMs. Models like CodeRetriever \cite{li-etal-2022-coderetriever} and CodeT5+ \cite{wang-etal-2023-codet5}, trained on large datasets and using advanced architectures, have significantly improved in generating source code. They use large-scale contrastive pre-training, making source code retrieval and generation more efficient and accurate.

While LLMs excel at generating natural language, they face challenges in many SE tasks, such as source code clustering or code-to-code search, which often require specialized training or fine-tuning to produce embeddings. Researchers are exploring the potential of LLMs for this purpose. 
Studies have shown that LLMs are strong text encoders, capable of producing high-quality embeddings that capture textual meaning and relationships \cite{zhang2024code}.
For instance, Khajezade et al. \cite{llm_code_clone_2024} used prompt templates to ask LLMs directly whether two code snippets are clones, thereby avoiding intermediate source code embeddings. However, this method sometimes yields incorrect answers due to LLMs' fixed context length, which limits their ability to process long source code snippets and can lead to erroneous responses. 
Despite their impressive zero-shot learning capabilities, LLMs face significant challenges in downstream SE tasks. First, each LLM has a fixed context length, limiting the amount of input it can process at once. This is problematic for tasks like source code clustering and code-to-code search, which require processing large source code fragments. When the input exceeds the context length, LLMs may produce error messages or malfunction. Second, LLMs often produce incorrect responses on complex tasks such as code clone detection. For example, an LLM might incorrectly assess the similarity between two source code fragments, determining that they have similar functionality but mistakenly responding ``no'' to indicate it did not recognize the samples as clone pairs. This issue undermines the reliability of LLMs in accurately identifying source code clones. Further details on these challenges are provided in Section~\ref{sec_motivation}.

To address issues in using LLMs for SE tasks, it is beneficial to convert source code fragments into embeddings. Source code embeddings are vector spaces that convert source code into meaningful vectors, which can be used for various downstream tasks, such as code summarization \cite{Sun2025, Tang2026} and vulnerability detection \cite{Haque2025}.
Guo et al. \cite{GUO2023103415} presented HetSum for code summarization via heterogeneous syntax graphs and dual position encoding.
Kamal et al. \cite{kamal2024interpretable} proposed an interpretable code summarization approach that leverages code embeddings to generate meaningful software descriptions, while Zhang et al. \cite{zhang2025boosting} introduced a context-based deep code representation method to enhance the identification of opportunities for identifier renaming in software projects. 
Some existing methods use LLMs to generate source code embeddings by leveraging their zero-shot learning capabilities, but they often require extensive training or fine-tuning resources. 
These limitations affect a wide range of software engineering applications, including code-clone detection \cite{higo2002software7}, clustering \cite{kuhn2007semantic}, and semantic code search \cite{liu2023graphsearchnet}.
Such challenges are also consistent with broader concerns regarding the testing and reliability of AI-enabled software systems \cite{Yi2025,Cuige2026}.
Existing approaches for source code representation mainly rely on pre-trained models, self-supervised learning, or fine-tuning strategies to obtain code embeddings. 
Although these approaches achieve promising performance, they often require additional training resources or labeled data. 
In contrast, \textsc{LSem2Vec} explores a training-free alternative by leveraging LLMs as semantic extractors and sentence embedding models for generating source code embeddings.
To overcome these challenges, we introduce a simple yet effective approach, \textsc{LSem2Vec} (\textit{\underline{\textbf{L}}LM-extracted code \underline{\textbf{Sem}}antics \underline{\textbf{to}} \underline{\textbf{Vec}}tor embedding}). This approach uses the zero-shot learning capabilities of an LLM and a sentence embedding model to generate high-quality source code embeddings without any training or fine-tuning.
In this paper, we focus specifically on unsupervised source code embedding generation rather than code representation learning in general or supervised representation training.
\textsc{LSem2Vec} uses an LLM to create summaries of source code fragments, which are then converted into source code embeddings through a sentence embedding model. 
In the experiments, we instantiate this embedding framework on two representative downstream tasks: unsupervised code-clone detection and code clustering.
\textsc{LSem2Vec} addresses common issues in software engineering tasks, providing an efficient and resource-effective solution. By eliminating the need for extensive training data and fine-tuning, \textsc{LSem2Vec} enhances the accuracy and reliability of source code embeddings while significantly reducing computational overhead, making it practical for real-world applications.
Compared with other frameworks, our proposed approach has several significant advantages for learning source code embeddings:
\begin{itemize}
        \item \textbf{\textit{Advantage 1:}} 
              It is flexible to generate source code embeddings using any LLMs and sentence embedding models. 
              This flexibility makes our method versatile for downstream tasks that require source code embeddings; in this paper, we concretely study source code clone detection and unsupervised clustering of source code fragments.
              For instance, we can apply model selection and a Gaussian-based mixture model \cite{xian_2021_AGGM,xian_2021_MML,Xian2022} to cluster source code fragments using the embeddings generated by \textsc{LSem2Vec}, without the need for any labels.

        \item
              \textbf{\textit{Advantage 2:}} 
              It does not depend on prior training or fine-tuning, making it inherently scalable across various programming languages. Unlike traditional approaches that often require large model sizes or extensive training data, our technique can generate source code embeddings directly from the code. This eliminates the need for any preliminary training or fine-tuning procedures. Such efficiency is particularly advantageous for real-world applications with limited computational resources and labeled data.
        \item
              \textbf{\textit{Advantage 3:}} 
              It employs a late-interaction approach to efficiently convert source code fragments into embeddings for code-clone detection, significantly reducing both time complexity and the number of LLM calls required. This method improves the efficiency of tasks involving source-code pairs, such as code-clone detection. Unlike traditional LLM-based methods that require examining each source-code pair individually, \textsc{LSem2Vec} leverages LLMs' zero-shot capabilities to convert each source-code fragment into embeddings. These embeddings are then utilized for further interaction computations, thereby reducing the number of LLM calls and the length of the LLM's input context.
        \item 
              \textbf{\textit{Advantage 4:}} 
              It mitigates the issue of generating erroneous responses in LLMs by breaking down the task into smaller, more manageable steps. A motivating example of erroneous response generation in LLMs is presented in Section~\ref{sec_error_resp}. \textsc{LSem2Vec} employs LLMs solely to extract code semantics, which are then transformed into embeddings for downstream SE tasks, with the embeddings introduced as a form of late interaction. By limiting the application of LLMs to source-code semantics extraction, we reduce the potential for errors and enhance the reliability of the generated embeddings. 
\end{itemize}

This paper introduces an approach for unsupervised source code embedding generation from unlabeled datasets, and evaluates the resulting embeddings on code-clone detection and code clustering.
To the best of our knowledge, this is the first work to employ large language and sentence embedding models to generate source code embeddings without requiring any training or fine-tuning.
Our main contributions to this paper are listed as follows:
\begin{itemize}
        \item[(1)]
              We first introduce a simple yet effective approach, \textsc{LSem2Vec}, which integrates large language and sentence embedding models to generate source code embeddings from unlabeled source code without requiring any training or fine-tuning.
              This approach is designed to be LLM-independent, allowing the use of any LLM, whether open-source or proprietary.
        \item[(2)]
              We propose a comprehensive pipeline to relieve two significant issues associated with LLMs: 
              erroneous responses and context-length limitations. 
              \textsc{LSem2Vec} effectively mitigates erroneous responses, ensuring the generated content is accurate and reliable. 
              Additionally, it mitigates the constraints imposed by LLMs' limited context length. 
    
        \item[(3)]
              \textsc{LSem2Vec} demonstrates strong performance on SE tasks. 
              We evaluate its effectiveness on the two studied source code-related tasks under different configurations, and demonstrate its superiority over existing methods such as SourcererCC \cite{sajnani2016sourcerercc},
              Code2vec \cite{alon2019code2vec,kang2019assessing}, InferCode \cite{buiInferCodeSelfSupervisedLearning2021}, TransformCode \cite{xian_2024}, and LLM2Vec \cite{behnamghader2024llm2vec}.
        \item[(4)] 
              \textsc{LSem2Vec} pioneers the integration of LLMs to generate high-quality source code embeddings. 
              These embeddings are shown to be effective for the evaluated downstream code-clone detection and code clustering tasks and may be extended to other source-code analysis tasks.
              By leveraging LLM capabilities, our method offers a novel direction for the community to explore and utilize advanced source code representations.
\end{itemize}

The remainder of this paper is structured as follows:
Section \ref{sec_motivation} gives some motivating examples for \textsc{LSem2Vec}.
Section \ref{sec_framework} provides our proposed approach for source code embedding. 
Section \ref{sec_experiment} presents the experimental design, including research questions, downstream tasks, datasets, and independent and dependent variables. 
Section \ref{sec_experiment_result} describes and discusses the experimental results.
Section \ref{sec:threatstovalidity} discusses threats to validity.
Finally, Section \ref{sec_conclusion} concludes this paper and outlines some potential future work.

\begin{figure}[!b]
\begin{tcolorbox}
\centering
 \begin{minipage}{0.95\textwidth}
\begin{lstlisting}[numbers=none, xleftmargin=0pt]
 {source code1}, {source code2},
 Do source code 1 and source code 2 solve identical problems with the same inputs and outputs? Answer with yes or no and no explanation.
\end{lstlisting}
\end{minipage}
\end{tcolorbox}
\caption{An improved prompt for code-clone detection}
\label{old_prompt}
\end{figure}

\begin{figure*}[!b]
\begin{tcolorbox}
\begin{minipage}[t]{0.5\textwidth}
        \begin{lstlisting}[language=c, caption=A sample of C code 1., label=fig_case_code_c_example1]
int main() {
  int n;
  int a[100000];
  int b[100000];
  cin >> n;
  for (int t = 0; t < n; t++) {
    a[t] = 0;
    b[t] = 0;
  }
  int i, j;
  while (cin >> i) {
    cin >> j;
    if (i == 0 && j == 0)
      break;
    else {
      a[i]++;
      b[j]++;
    }
  }
  for (int r = 0; r < n; r++) {
    if (a[r] == 0 && b[r] == n - 1) {
      cout << r << endl;
    }
  }
  return 0;
}\end{lstlisting}
\end{minipage}
    \begin{minipage}[t]{0.5\textwidth}
        \begin{lstlisting}[language=c, caption=A sample of C code 2., label=fig_case_code_c_example2]
int ren[1000000][2], ming[1000000][2];
int main() {
  int n, i = 0, num = 0;
  memset(ming, 0, sizeof(ming));
  cin >> n;
  while (1) {
    cin >> ren[i][0] >> ren[i][1];
    if (ren[i][0] == 0 && ren[i][1] == 0)
      break;
    else {
      ming[ren[i][0]][0]++;
      ming[ren[i][1]][1]++;
    }
    i++;
  }
  for (i = 0; i < n; i++) {
    if (ming[i][0] == 0 && ming[i][1] == n - 1) {
      cout << i << endl;
      num++;
    }
  }
  if (num == 0)
    cout << "NOT FOUND" << endl;
  return 0;
}\end{lstlisting}
\end{minipage}
\end{tcolorbox}
        \caption{An example of code-clone in C}
        \label{fig_case_c_code}
\end{figure*}

\section{Motivating Examples}
\label{sec_motivation}
This section provides two motivating examples for our method:  \textit{erroneous responses} and \textit{late interaction}.

\subsection{Erroneous Responses in LLM}
\label{sec_error_resp}

LLMs often struggle to produce erroneous responses, which can lead to incorrect outputs and unreliable results \cite{yao2023llm}. This issue is particularly evident in scenarios where LLMs are tasked with identifying whether two pieces of source code are clones. Although LLMs might possess a comprehensive understanding of the code's functionalities, they frequently fail to accurately determine source-code cloning \cite{llm_code_clone_2024}. To address this challenge, Khajezade et al. \cite{llm_code_clone_2024} proposed an enhanced prompting strategy. Rather than directly querying whether the codes are clones, they advocated for a more precise inquiry: asking if the two codes yield identical inputs and outputs. This nuanced approach aims to reduce the incidence of erroneous responses by aligning LLM evaluation with the code's functional equivalence, as illustrated in Figure~\ref{old_prompt}.
This approach constrains LLMs to respond with either ``Yes'' for clone pairs or ``No'' for non-clone pairs, thereby enhancing response accuracy.

We examine two source-code samples from the POJ-104  dataset \cite{zhangNovelNeuralSource2019,mouConvolutionalNeuralNetworks2015}, as illustrated in Figure~\ref{fig_case_c_code}. 
Using these samples as a case study, we prompted GPT-3.5 with a custom template to determine whether they are clone pairs. 
Consistently, the model responded with ``no'' indicating that it did not recognize the samples as clone pairs. 
However, according to the dataset, these samples share label 85, indicating they are indeed clone pairs. 
This discrepancy highlights a significant issue: Despite employing a newly designed prompt, LLMs like GPT-3.5 turbo can still produce erroneous responses.
This case study underscores the need for further refinement of the workflow for using LLMs in software engineering tasks such as code clone detection.

\begin{figure}[!t]
\begin{tcolorbox}
\centering
 \begin{minipage}{\textwidth}
 \small
\begin{lstlisting}
{
  "error": {
    "message": "This model's maximum context length is 4096 tokens. However, you requested 4105 tokens (4008 in the messages, 97 in the completion). Please reduce the length of the messages or completion.", "type": "invalid_request_error", "param": "messages", "code": "context_length_exceeded"
  }
}
\end{lstlisting}
\end{minipage}
\end{tcolorbox}
\caption{Error message from GPT-3.5}
\label{fig_out_of_limited}
\end{figure}

\begin{figure*}[!b]
        \centering
        \graphicspath{{materials_img/}}
                \includegraphics[width=\textwidth]{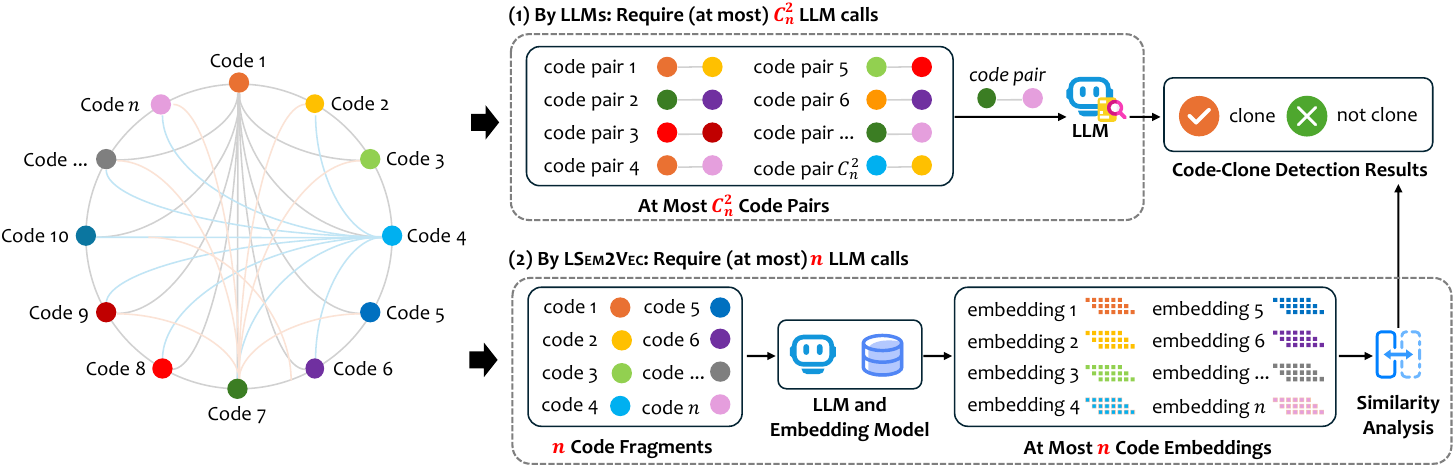}
\caption{
Side-by-side comparison of two code clone detection workflows to illustrate the efficiency advantage of the late-interaction design in LSem2Vec.
(1) Upper subgraph: The naive direct LLM prompting paradigm. For all $C_n^2$ code pairs, each pair must be concatenated and fed into LLMs separately, leading to quadratic LLM invocation overhead and frequent context-length overflow for long code snippets.
(2) Lower subgraph: Our proposed LSem2Vec pipeline. Each code fragment only triggers one LLM call to generate a semantic summary, then sentence embeddings are precomputed and cached. Pairwise similarity calculation is conducted offline on embedding vectors without extra LLM requests, reducing time complexity from $O(n^2)$ to $O(n)$.
(3) This figure motivates the late-interaction design to resolve context limit and high token cost issues.}
        \label{fig_code_clone}
\end{figure*}

The above case study shows that erroneous responses from LLMs often stem from complex workflows. 
For instance, in code clone detection, LLMs are tasked with summarizing the functionality of two source code pairs and then determining whether they are clones. 
Despite LLMs' ability to accurately summarize the functionality of source-code pairs, the extended and complex nature of this workflow can lead to erroneous results. 
These erroneous responses occur when the model generates outputs that deviate from the intended meaning or factual accuracy. 

We use the term ``erroneous responses'' to describe outputs generated by LLMs that are factually incorrect, logically inconsistent, or unsuitable for the intended software engineering task. Such errors often arise when the model is forced to handle inputs exceeding its context length or when complex tasks are attempted without structured decomposition. In our framework, erroneous responses are mitigated by breaking tasks into smaller steps—summarizing code fragments, generating embeddings, and performing downstream analysis—thereby reducing the risk of inaccuracies.

\subsection{Late Interaction} 
\label{sec_late_interaction}

While LLMs have significantly advanced the field of software engineering, they still encounter notable limitations in specific tasks due to their restricted context length. The context length of an LLM is fixed and limited, and increasing it requires substantial GPU resources for retraining or fine-tuning. Context length refers to the maximum number of tokens (words, punctuation, etc.) the model can process at once, encompassing both the input and output of the LLMs.

For instance, the standard GPT-3.5 has a maximum context length of 4,096 tokens. This limitation is insufficient to complete software engineering tasks such as classification, clustering, and code search. 
As illustrated in Figure~\ref{fig_case_c_code}, processing a single source code fragment can consume a significant portion of the context length (Using GPT-3.5 as an example): Listing \ref{fig_case_code_c_example1} requires 164 tokens, while Listing \ref{fig_case_code_c_example2} requires 202 tokens. 
Given an average token count of 160 per source code fragment, GPT-3.5 can handle only approximately 25 source code fragments simultaneously, which is inadequate for source-code datasets.
Even with the extended GPT-3.5 16k, which has a maximum context length of 16,385 tokens, the model can process only about 100 source code fragments, given an average token count per fragment of 160. 
In this paper, we utilize the standard GPT-3.5 with a maximum context length of 4,096 tokens. 
An example of an error message encountered due to exceeding the context length in GPT-3.5 is shown in Figure~\ref{fig_out_of_limited}.

Similarly, in code-to-code search, the limited context length poses a challenge, hindering the model's ability to effectively match and retrieve relevant source code segments. SE tasks often require processing and analyzing large volumes of source code fragments, which are constrained by limited context length.

Another significant concern is the high cost associated with the tokens required for LLMs in SE tasks. For instance, as illustrated in Figure \ref{fig_code_clone}, in the task of source code clone detection, there are $n$ unique source code fragments that need to be analyzed. These fragments can form $C_{n}^{2}$ possible pairs, each of which requires a separate LLM call for evaluation. Consequently, using LLMs for this task requires a huge number of token-based computations, which is very expensive and inefficient. Additionally, the time and computational resources needed to handle so many calls make it hard to use LLMs effectively for SE tasks.

In the previous examples of SE tasks, we noted that using LLMs to analyze a large number of source code snippets simultaneously is neither practical nor scalable. This challenge is particularly evident in code-clone detection, where LLMs must analyze each code pair individually without the benefit of caching, making the process inefficient. Inspired by these examples, we propose a late interaction method for SE tasks. Instead of using LLMs to directly find interactions between source code snippets in a single LLM call, we suggest using LLMs to extract the semantic meaning of the source code and convert it into embeddings. These embeddings can then be used later for calculations, a process we refer to as ``late interaction'' in this paper.

However, current LLMs are not designed to generate source code embeddings directly; they produce natural language output, which is unsuitable for the tasks we discussed. Additionally, when used for complex tasks, LLMs often produce incorrect results. To overcome these issues, we propose breaking down the workflow into smaller, more manageable steps. Specifically, we primarily use an LLM to summarize source code and then apply an embedding model to convert these summaries into vector representations. As a result shown in Figure \ref{fig_code_clone}, this approach significantly improves the efficiency of code-clone detection, reducing the complexity from $O(n^2)$ with traditional LLM methods to $O(n)$.

This approach aims to improve the accuracy and reliability of SE tasks by reducing the risk of incorrect results through a more structured and step-by-step process.

In this paper, ``late interaction'' refers to a strategy for minimizing LLM calls in software engineering tasks. Instead of requiring the LLM to directly process and compare multiple source code fragments within a single context window, we first use the LLM to extract semantic meaning from individual fragments and convert them into embeddings. These embeddings are then stored and later used for similarity calculations or clustering. By deferring the interaction between code fragments to the embedding space, this approach reduces computational overhead, avoids repeated LLM queries, and scales more efficiently to large datasets.

\section{Approach}
\label{sec_framework}

\begin{figure*}[!t]
        \centering
        \graphicspath{{materials_img/}}

                \includegraphics[width=\textwidth]{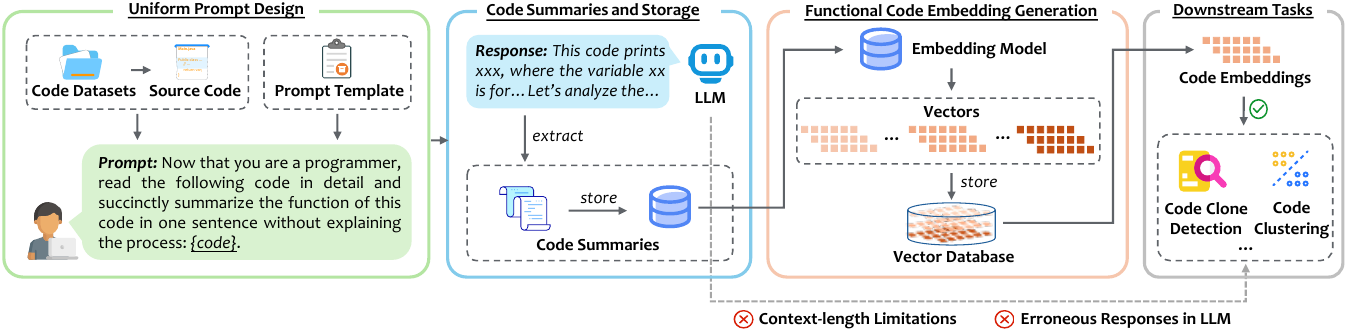}
\caption{
Full four-stage pipeline of the proposed \textsc{LSem2Vec} approach for unsupervised source code embedding generation.
(1) Pipeline layout description: The diagram is horizontally divided into four independent sequential modules separated by blank spacing, connected by solid arrows for data transmission and dashed arrows for local disk storage. Module 1: Uniform one-sentence prompt template construction for cross-language LLM code summarization; Module 2: LLM single-sentence summary generation and persistent storage of semantic intermediate representations; Module 3: Sentence embedding model inference to convert text summaries into fixed-dimensional semantic vectors; Module 4: Reusable embeddings for two core downstream tasks: code clone detection and unsupervised code clustering.
(2) Method implication: This late-interaction pipeline only invokes LLMs once per code snippet, avoiding repeated pairwise LLM calls required by direct LLM prompt-based clone detection.}
        \label{fig_proposed}
\end{figure*}

In this section, we provide the \textsc{LSem2Vec} approach for source code embedding that leverages zero-shot learning with LLMs.
Figure \ref{fig_proposed} presents the framework of \textsc{LSem2Vec}, which mainly consists of four manageable steps:
(1) uniform prompt design;
(2) source-code summaries and storage;
(3) source code embedding generation; and support for
(4) downstream tasks.
The two-stage design of \textsc{LSem2Vec} is motivated by the complementary capabilities of LLMs and sentence embedding models. 
The LLM-based summarization stage extracts high-level semantic information from source code, while the sentence embedding stage converts the generated semantic descriptions into fixed-dimensional vectors. 
Therefore, LLMs are used as semantic extractors rather than direct embedding generators, which enables \textsc{LSem2Vec} to leverage LLMs without requiring additional training or fine-tuning.
For ease of description, the term ``\textit{code}'' represents ``\textit{source code}'' in the rest of this study, unless explicitly stated.

\subsection{Uniform Prompt Design}
\label{subsubsec_prompt}
In this step, our objective is to propose a standardized English-language prompt template that can subsequently be translated into various languages to ensure its applicability across diverse sentence models. This approach aims to facilitate the creation of prompts that are both consistent and adaptable for different language models. However, there are two crucial considerations when designing the prompt template:
(1) summarize the function of the code into one sentence; and 
(2) avoid LLMs from providing any explanatory content. 
The first aspect arises from the limitations of sentence embedding models used in \textsc{LSem2Vec}: 
Sentence embedding models are pre-trained on sentence pairs, typically formatted as single sentences. 
Consequently, LLMs must generate one-sentence code summaries to maintain compatibility with the sentence embedding models.
The second aspect stems from the fact that LLMs tend to provide excessive detail at the beginning or end of their responses. 
Furthermore, we avoid incorporating task-specific instructions into the prompt because the generated summaries are designed as general semantic representations of source code. 
The same prompt structure is therefore applied to different downstream tasks to ensure that the generated summaries are comparable.
This phenomenon can introduce variability and uncertainty into \textsc{LSem2Vec}. 
To mitigate this risk, we instruct LLMs to refrain from any explanations. 
However, some LLMs may still include detailed explanations, and the code summary is consistently presented in the first sentence.
To address this issue, we obtain a reliable summary only from the first sentence of the response for each code fragment.

The flexibility of \textsc{LSem2Vec} allows us to extend to various languages, accommodating embedding models trained in specific languages.
When using different language-based prompts for LLMs, the generated summaries will vary depending on the language used.
For example, Figure~\ref{fig_prompt} presents an example template of the designed prompt in English.
For all LLM-based semantic extraction, we used the same prompt structure across models and kept the decoding configuration, including temperature and related generation settings, fixed throughout the experiments. 
The English and Chinese summary settings differ only in the required output language; the intended semantic content, including functionality, inputs, outputs, and key operations, remains the same. 
Unless otherwise stated, no task-specific prompt tuning is performed. 
The generated summaries are then tokenized by the corresponding sentence embedding model, and the stop-word removal setting is only used in the ablation study.

The summaries are intended to preserve semantic information needed for downstream embedding, including the code purpose, inputs and outputs, important API or library operations, and key control-flow behavior. 
Function names, variable names, and other identifiers are retained when they carry semantic meaning; otherwise, the summary emphasizes their roles rather than only their surface forms.

\begin{figure}[!h]
    \centering
    \begin{tcolorbox}
        {\small Now that you are a programmer, read the following code in detail and succinctly summarize the function of this code in one sentence without explaining the process:\\
        \texttt{code}}
    \end{tcolorbox}
    \caption{Example templates of the designed prompt}
    \label{fig_prompt}
\end{figure}

\begin{figure*}[!t]
\begin{tcolorbox}
A sample of C code:\\
\begin{minipage}{0.48\textwidth}
        \begin{lstlisting}[language=c, caption=, label=fig_case_code_c_example]

int n, a[15];

double ck(int x) {

  if (x >= 90) return 4.;
  if (x >= 85) return 3.7;
  if (x >= 82) return 3.3;
  if (x >= 78) return 3.;
  if (x >= 75) return 2.7;
  if (x >= 72) return 2.3;
  if (x >= 68) return 2.;
  if (x >= 64) return 1.5;
  if (x >= 60) return 1.;
  
  return 0;

}
\end{lstlisting}
\end{minipage}
 \hfill
\begin{minipage}{0.48\textwidth}
\begin{lstlisting}[language=c, caption=, label=fig_case_code_c_example,firstnumber=19]
int main() {
  while (scanf("%d", & n) == 1) {
    int sum1 = 0;
    double sum2 = 0;
    for (int i = 0; i < n; i++) {
      scanf("%d", a + i);
      sum1 += a[i];
    }
    for (int i = 0; i < n; i++) {
      int tmp;
      scanf("%d", & tmp);
      sum2 += ck(tmp) * a[i];
    }

    printf("%.2f", sum2 / sum1);
  }
  return 0;
}
\end{lstlisting}
\end{minipage}
\begin{minipage}{0.98\textwidth}
\begin{tcolorbox}[breakable,fontupper=\sffamily\bfseries,colback=gray!0,colframe=black,arc=1mm,left={1mm},top={1mm},bottom={1mm},right={1mm},boxrule={0.05mm}]
        {\small \textit{\textbf{Code summary:}} \\
         This code calculates and prints the weighted average of a set of grades, where the weights are determined by the corresponding credit hours, and the grades are converted to a GPA scale using the `ck' function}
\end{tcolorbox}
\end{minipage}
\end{tcolorbox}
        \caption{A sample of C code and its summary}
        \label{fig_case_code_sum}
\end{figure*}

\subsection{Code Summaries and Storage}
\label{sec-code-summaries and storage}

In this step, \textsc{LSem2Vec} focuses on using LLMs to generate concise code summaries based on previously designed prompts. Our method involves extracting the first sentence from the LLMs' responses and using it as the code summary. It's essential to note that LLMs will respond in the same language as the prompts. Once the code summaries are generated, they are stored as intermediate semantic representations of the code, which can be used in subsequent steps. For convenience, these code summaries are stored in the JSON format, although other storage formats can also be used. Figure~\ref{fig_case_code_sum} shows an example of C code and its corresponding code summary generated by GPT-3.5.

In code-clone detection tasks, traditional LLM-based approaches must combine two code fragments as input to LLM prompts. If two code fragments are lengthy, the input may exceed the context-length limits of each LLM session, resulting in incomplete or incorrect responses.
\textsc{LSem2Vec} can effectively address these two issues by summarizing each long code fragment individually, avoiding simultaneous submission to the LLM, thereby providing more accurate and reliable code summaries for downstream tasks.

Moreover, traditional LLM-based approaches often necessitate adjusting the prompt template to suit specific downstream tasks, even when dealing with the same dataset. Such frequent modifications can be time-consuming and resource-intensive. \textsc{LSem2Vec} introduces a storage mechanism that captures and stores extracted code summaries along with the corresponding code fragments of each dataset. In \textsc{LSem2Vec}, we persist all code summaries to disk in the JSON format for each dataset. Specifically, by generating code summaries only once per dataset, we eliminate the need for repeated computations. All LLM-generated code summaries are stored, allowing them to be utilized in various downstream tasks.

\subsection{Code Embedding Generation}

In this step, our primary objective is to generate code embeddings by converting code summaries obtained from the previous stage into vector representations. This conversion is achieved using a language-specific sentence embedding model. These pre-trained sentence embedding models effectively transform code summaries into embeddings that encapsulate the semantic information inherent in the code.

To ensure consistency, our prompt design enables LLMs to generate one-sentence code summaries. These concise summaries are then efficiently converted into code embeddings by a sentence embedding model. It is crucial to highlight that these models only support code summaries in the same language. For instance, certain sentence embedding models are pre-trained exclusively in English, thereby restricting their applicability to English code summaries.

Moreover, the uniform prompt design not only standardizes the code summary generation process but also enhances the model's ability to produce high-quality embeddings. This consistency is vital for accurate representation and subsequent analysis of the code. It is essential to note that the choice of sentence embedding models directly impacts the quality and applicability of the generated embeddings. As such, selecting the appropriate model is a critical consideration in this process.

Furthermore, the flexibility of \textsc{LSem2Vec} allows us to integrate different sentence embedding models, accommodating future advancements and ensuring our method remains adaptable to evolving technologies. This adaptability is particularly important, as it enables the incorporation of newer, more sophisticated models, thereby enhancing the robustness and effectiveness of our code-embedding generation framework.

\subsection{Downstream Tasks}

The framework of \textsc{LSem2Vec} can generate code embeddings without training or labeled data, making it an extremely valuable tool for a range of downstream tasks in software engineering. 
In this paper, we specifically illustrate how \textsc{LSem2Vec} can be effectively applied to unsupervised code-clone detection across multiple programming-language datasets and to code clustering.
\textsc{LSem2Vec} leverages a pre-trained LLM and a sentence-embedding model to perform the aforementioned software-engineering tasks. The results from our experiments demonstrate that \textsc{LSem2Vec} significantly outperforms all existing unsupervised learning methods.

Compared with traditional LLM-based approaches, our method demonstrates greater efficiency and cost-effectiveness, owing to fewer LLM calls and reduced token usage. Unlike existing PTMs, \textsc{LSem2Vec} eliminates the need for training or fine-tuning, further enhancing efficiency. This paper will delve deeper into these advantages in the subsequent experimental section, providing detailed insights into the performance metrics.

\section{Experimental Design}
  \label{sec_experiment}
In this section, we first present the research questions related to the performance of \textsc{LSem2Vec} and then
formulate the tasks we conducted to address them from the perspectives of code clone detection and code clustering.
We also outline the datasets and the independent and dependent variables used in our experiments.
Additionally, we briefly introduce the experimental environment for our research.

\subsection{Research Questions}
\label{sec-rqs}
To thoroughly evaluate the effectiveness of \textsc{LSem2Vec}, we aim to answer the following research questions in the following experiments:
\begin{description}[leftmargin=0pt] 

\item[\textbf{RQ1:}] \textbf{[Ablation Study]} 
\begin{itemize}[leftmargin=20pt, labelindent=0pt, itemindent=0pt]
    \item \textbf{RQ1.1:} What is the impact of using different sentence embedding models trained in different languages on the effectiveness of \textsc{LSem2Vec}? 
    \item \textbf{RQ1.2:} What is the impact of removing stop words from the LLM's response on the effectiveness of \textsc{LSem2Vec}? 
    \item \textbf{RQ1.3:} What is the impact of using different LLMs for \textsc{LSem2Vec}? 
    \item \textbf{RQ1.4:} What is the performance difference between our proposed approach and the implementation of code clone detection using a prompt within the LLM utilized in \textsc{LSem2Vec}?
\end{itemize}

\item[\textbf{RQ2:}] \textbf{[Generalization Evaluation]} Does \textsc{LSem2Vec} also support the generation of code embeddings for code fragments in other programming languages?

\item[\textbf{RQ3:}] \textbf{[Effectiveness Evaluation]} How does the effectiveness of \textsc{LSem2Vec} compare to other unsupervised code embedding approaches? 
\item[\textbf{RQ4:}] \textbf{[Quality Evaluation]} How effective are the code embeddings generated by \textsc{LSem2Vec} when evaluated through visualization techniques, especially regarding boundary separation effects across various LLM configurations? 
\end{description}

We first conduct a comprehensive series of ablation experiments to identify the optimal configuration of \textsc{LSem2Vec} for each LLM. It should be noted that we also conducted an additional version that excludes stop words from the Chinese code summaries for \textsc{LSem2Vec}. This version is specifically designed to evaluate the impact of stop words on the performance of embedding models (more details will be discussed in Section \ref{sec_experiment_result}). 
These ablation experiments are meticulously designed from various perspectives:
\textbf{RQ1.1} and \textbf{RQ1.3} compare the impact on the effectiveness of using different sentence embedding models and LLMs, respectively. 
\textbf{RQ1.2} aims at evaluating the impact of stop words on the performance of sentence embedding models.
By comparing the performance of models with and without stop words, we can gain insights into the role of these words in embedding quality and overall model effectiveness. 
\textbf{RQ1.4} seeks to evaluate and compare the effectiveness of two code clone detection approaches: \textsc{LSem2Vec} and the baseline LLMs utilized within it.
\textbf{RQ2} and \textbf{RQ3} respectively evaluates the generalization and effectiveness of \textsc{LSem2Vec}.
\textbf{RQ4} evaluates the quality of \textsc{LSem2Vec} from the visualization perspective.

\subsection{Task Formulation}
\label{SEC: Task Formulation}

We formulate two unsupervised downstream tasks to evaluate the performance of \textsc{LSem2Vec}: 
code-clone detection; and 
code clustering.

\subsubsection{Code-Clone Detection}
\label{sec-code-clone-detection-task}
Code-clone detection involves identifying code fragments that are similar or identical in syntax or semantics. 
The code-clone detection is essential for maintaining code quality, reducing redundancy, and facilitating software maintenance \cite{higo2002software7}. 
According to the previous study \cite{liuCanNeuralClone2021}, code clones can be categorized into four major types:
\begin{itemize}
    \item
    \textbf{Type 1:} 
    \textit{Exact copies} refers to identical code fragments except for variations in white space, layout, and comments. 
    These are also known as exact or textual clones.

    \item
    \textbf{Type 2:}
    \textit{Syntactically similar but with variations} refers to identical code fragments except for variations in identifier names and literal values.
    These are also known as renamed or parameterized clones.

    \item
    \textbf{Type 3:}
    \textit{Copied with further modifications} refers to the code fragments that are syntactically similar but different at the statement level.
    These are also known as gapped or near-miss clones.

    \item
    \textbf{Type 4:} 
    \textit{Semantically similar but syntactically different} refers to the code fragments that are syntactically different but implement the same functionality.
    These are also known as semantic or functional clones.
\end{itemize}

In this paper, we address the challenge of detecting type 4 code clones, which are particularly difficult to identify, using an unsupervised approach. These clones are semantically similar but differ syntactically, meaning that they perform the same functions but have different code structures. 
The task of unsupervised code clone detection for our method can be defined as follows: There is a set of source-code files, \( \{C_1, C_2, ..., C_n\} \). Given a set of code clones, \( \{CP_1, CP_2, ..., CP_m\} \), where each clone \( CP_i \) is a pair of code fragments \( (F_i, G_i) \) such that \( F_i \subseteq C_j \) and \( G_i \subseteq C_k \) for some \( j, k \in \{1, 2, ..., n\} \).
We define \( S(F,G) \) as a similarity score, where a larger value indicates a more semantically similar pair. A code pair \( (F_i,G_i) \) is predicted as a clone when \(S(F_i,G_i)\) reaches or exceeds the predefined threshold \(T\).
The task can then be formulated as:
\begin{equation}
\begin{aligned}
 \{(F_i, G_i)\} = &\{(F, G) \mid S(F, G) \geq T  \text{ for all } F \subseteq C_j \\
&\text{ and } G \subseteq C_k \text{ with } j, k \in \{1, 2, ..., n\}\}, 
\end{aligned}
\end{equation}
where \( T \) is the similarity threshold.
Because \(S\) denotes similarity rather than distance in clone detection, increasing \(S\) increases the likelihood of predicting a clone.

\subsubsection{Code Clustering}
\label{sec-code-clustring-task}
Code clustering involves automatically grouping similar code fragments into clusters without supervision, a crucial task for various applications in software engineering and code analysis \cite{kuhn2007semantic}. 
However, due to limited context length and the sheer volume of code to process simultaneously, it is impractical to use LLMs directly for this task.
Code clustering typically involves transforming code fragments into numerical representations, known as embeddings, that capture the code's semantic attributes. 
The following is its formal definition using a distance-based criterion for clustering:
\begin{enumerate}
    \item \textbf{Code Embedding Generation}: Convert each code fragment $C_i$ into a code embedding $E_i$.
    \item 
    \textbf{Distance Calculation}: Define a distance metric \(D(E_i, E_j)\) between embeddings \(E_i\) and \(E_j\), where smaller values indicate higher similarity.
    \item \textbf{Clustering Algorithm}: Apply a clustering algorithm \cite{xian_2021_MML,xian_2021_AGGM,Xian2022,kmeans_2002} to group similar embeddings into clusters.
\end{enumerate}


For effective clustering, we use Euclidean distance between embeddings rather than a similarity score; smaller distances indicate more similar code fragments. The distance is defined as follows:
\begin{equation}
{ D(E_i, E_j) = \|E_i - E_j\|.}
\end{equation}
Euclidean distance is adopted because it is consistent with the objective function of K-means, which minimizes the intra-cluster variance based on Euclidean distance.
Besides, we employ $K$-means \cite{kmeans_2002}, a widely used clustering algorithm \cite{xian_2021_MML,xian_2021_AGGM,Xian2022}, to organize the code fragments into meaningful clusters. 
The $K$-means algorithm iteratively partitions the data into $K$ clusters by minimizing the variance within each cluster, thereby grouping similar code fragments. Its formula is defined as follows:
\begin{equation}
\arg \min_{\{C_1, C_2, \ldots, C_K\}} \sum_{k=1}^K \sum_{E_i \in C_k} \|E_i - \mu_k\|^2, 
\end{equation}
where \(\mu_k\) is the mean of the embeddings in cluster \(C_k\).

\begin{table*}[!t]
 \centering
 \caption{Dataset summary}
 \label{TAB:dataset}
 \setlength\tabcolsep{1.2mm}
 \small
 \begin{tabular}{clcrcrrrc}
 \hline
 \textbf{No.} &\textbf{Name} &\textbf{Language} & \textbf{Samples} & \textbf{Pair Format} & \textbf{Train Samples} &\textbf{Validation Samples} & \textbf{Test Samples} & \textbf{Year} \\ \hline
 1 & POJ-104 \cite{zhangNovelNeuralSource2019,mouConvolutionalNeuralNetworks2015}& C & 52,000 & N & 32,000 & 8,000 & 12,000 & 2016  \\
 2 & OJClone C \cite{mouConvolutionalNeuralNetworks2015,xian_2024} & C & 10,000 & Y & 7,000 & 1,000 & 2,000 & 2024  \\
 3 & BigCloneBench \cite{svajlenko2014towards, wang2020detecting}  & Java & 1,731,860 & Y & 901,028 & 415,416 & 415,416 & 2014 \\
 \hline
 \multicolumn{2}{c}{\textbf{\textit{Sum}}} &-- & 1,793,860 &-- &{940,028} &424,416 &429,416 &--\\\hline
 \end{tabular}
\end{table*}

\begin{table*}[!b]
 \centering
 \small
 \caption{Summary of selected LLMs}
 \label{TAB:llm-baselines}
 \setlength\tabcolsep{3.7mm}
 \begin{tabular}{cllcccclc}
 \hline
 \textbf{No.} &\textbf{Model Name} &\textbf{Version} & \textbf{Architecture} & \textbf{Parameter} & \textbf{Organization}  & \textbf{Year} \\ \hline
 1 & GPT-3.5 \cite{GPT35} & gpt-3.5-turbo & Decoder-only & 175B & OpenAI  & 2023 \\
 2 & GPT-5.4 \cite{GPT54}  & gpt-5.4-turbo & Decoder-only & Unknown & OpenAI  & 2026 \\
 3 & GLM3~\cite{glm} & glm-3 & Decoder-only & 6B & Zhipuai  & 2023 \\
 4 & GLM4~\cite{glm} & glm-4 & Not Reported & 9B & Zhipuai  & 2024 \\
 \hline
 \end{tabular}
\end{table*}

\begin{table*}[!t]
 \centering
 \caption{Summary of sentence embedding models}
 \label{TAB:embedding_model}
 \setlength\tabcolsep{1.7mm}
 \small
 \begin{tabular}{cllccccc}
 \hline
 \textbf{No.} &\textbf{Model Name} &\textbf{Version} & \textbf{Type} & \textbf{Hidden Size}  & \textbf{Vocab Size} & \textbf{No. of Layers} & \textbf{Language} \\ \hline
 1 &MiniLM-L6 \cite{MiniLM_L6} & all-MiniLM-L6-v2 & BERT & 384  & 30,522 & 6 & English \\
 2 &MiniLM-L12 \cite{MiniLM_L12} & all-MiniLM-L12-v2 & BERT & 384  & 30,522 & 12 & English \\
 3 &SBERT-CN \cite{sbert_base_chinese} & sbert-base-chinese-nli & BERT & 768  & 21,128 & 12 & Chinese \\\hline
 \end{tabular}
\end{table*}

\subsection{Dataset Selection}
\label{sec_dataset}
For our experiments, we utilize two prominent datasets: POJ-104 \cite{zhangNovelNeuralSource2019,mouConvolutionalNeuralNetworks2015} and BigCloneBench \cite{svajlenko2014towards, wang2020detecting}.
POJ-104 is a dataset specifically designed for code-clone detection tasks. 
POJ-104 consists of 52,000 code fragments written in C, which are semantically equivalent but syntactically different \cite{zhangNovelNeuralSource2019}. 
This dataset is structured to facilitate the evaluation of models based on their ability to identify semantically similar code fragments despite syntactic variations. 
The dataset is divided into training, validation, and test sets, with 32,000, 8,000, and 12,000 examples, respectively.
BigCloneBench, a widely used benchmark dataset \cite{svajlenko2014towards, wang2020detecting}, includes projects from 25,000 Java repositories and covers 10 functionalities. 
It contains 6,000,000 true clone pairs and 260,000 false clone pairs. 
This extensive dataset comprehensively evaluates code-clone detection models, providing a robust benchmark for assessing their performance. 
Both datasets are available from the CodeXGLUE GitHub repository\footnote{\textbf{CodeXGLUE}: \url{https://github.com/microsoft/CodeXGLUE}.}.

Additionally, following prior research \cite{xian_2024}, we employ the OJClone C dataset, which contains code pairs from POJ-104, ranked by pairwise similarity.
This dataset comprises 500 programs from each of the first 15 POJ-104 problems, yielding 1.8 million clone pairs and 26.2 million non-clone pairs.
A comparison of all the pairs would be prohibitively time-consuming, so 5,000 clone pairs and 5,000 non-clone pairs were randomly selected for the code-clone detection evaluation. 
Note that OJClone C will be utilized for all samples in the unsupervised code-clone detection experiments, as none of the approaches in these experiments require labeled data.
The detailed comparison of the used datasets is presented in Table~\ref{TAB:dataset}.
The datasets support different parts of the experimental design: OJClone C is used for the C-language code-clone detection ablations in RQ1 and the clone-detection effectiveness comparison in RQ3; BigCloneBench is used to evaluate Java code-clone detection and cross-language generalization in RQ2; and POJ-104 supports code clustering and visualization-based embedding-quality analysis in RQ3 and RQ4. This mapping clarifies how each dataset contributes to the two evaluated downstream tasks.

\subsection{Independent Variables}
We focus on different LLMs, sentence embedding models, and baselines for code clustering tasks as the independent variables in our experimental research.

\subsubsection{LLM Selection}
We employ four distinct LLMs for further evaluating the effectiveness of \textsc{LSem2Vec}: GPT-3.5 \cite{GPT35}, GPT-5.4 \cite{GPT54}, GLM3 \cite{glm}, and GLM4 \cite{glm}.
These LLMs were selected to provide diverse capabilities and performance characteristics, allowing for a comprehensive evaluation of \textsc{LSem2Vec}.
Table~\ref{TAB:llm-baselines} presents detailed information about the three LLMs. 
GPT-3.5 and GPT-5.4 have been widely recognized as powerful LLMs, while GLM3 and GLM4 are parts of the open-source GLM series of LLMs. 
These LLMs provide diverse configurations in terms of model scale, architecture, and availability, enabling us to evaluate whether \textsc{LSem2Vec} can generalize across different LLM settings.
Our primary objective is to evaluate the performance of \textsc{LSem2Vec} across various LLM configurations. 
While this study examines these three LLMs, \textsc{LSem2Vec} is flexible and can be adapted to integrate other closed-source LLMs that may offer superior performance.
This adaptability ensures that \textsc{LSem2Vec} remains relevant and effective as new and more advanced LLMs become available.

\subsubsection{Sentence Embedding Model Selection} 
We utilize the MiniLM-L6 with the version all-MiniLM-L6-v2 \cite{MiniLM_L6} and the MiniLM-L12 with the version all-MiniLM-L12-v2 \cite{MiniLM_L12} models from sentence transformers (SBERT) \cite{reimers-gurevych-2019-sentence} for English code summaries. For Chinese code summaries, we employ the SBERT-CN with the version sbert-base-chinese-nli model \cite{sbert_base_chinese}.
This dual-language implementation ensures that the code summaries are accurately generated in the appropriate language and enhances the overall quality and relevance of the generated embeddings.
Table~\ref{TAB:embedding_model} summarizes these three sentence embedding models used in \textsc{LSem2Vec}.
The MiniLM-L12 and MiniLM-L6 models share the same architecture; the primary difference is the number of layers: MiniLM-L12 has 12 layers, whereas MiniLM-L6 has 6.
Both models produce output with a hidden size of 384. 
In contrast, the SBERT-CN model, although also based on the BERT architecture, has a hidden size of 768 and a vocabulary size of 21,128 tokens, which is smaller than the vocabulary sizes of MiniLM-L12 and MiniLM-L6. 
Overall, the selected sentence embedding models cover both English and Chinese scenarios with different model capacities, allowing us to investigate the impact of embedding models on the performance of \textsc{LSem2Vec}.
The SBERT-CN model also comprises 12 layers but is specifically trained on Chinese datasets, making it more suitable for Chinese-related tasks.
We use general-purpose sentence embedding models because \textsc{LSem2Vec} first converts source code into natural-language semantic summaries. At this second stage, the input to the embedding model is no longer raw code but a concise sentence, so sentence embedding models are a natural fit for capturing summary-level semantics without additional code-specific training or fine-tuning. Code-specific embedding models remain important baselines, and we compare against representative code-oriented methods in the downstream evaluations.

\subsubsection{Baseline Methods}
To provide a comprehensive evaluation of \textsc{LSem2Vec}, we compared it against a diverse set of baseline approaches widely used in software engineering tasks. 

For code-clone detection, we included several advanced unsupervised methods such as Deckard \cite{jiang2007deckard}, which represents code structure using abstract syntax trees, and SourcererCC \cite{sajnani2016sourcerercc}, which relies on token-based representations. DLC \cite{white2016deep} and Code2vec \cite{alon2019code2vec,kang2019assessing} apply deep learning techniques to embed code snippets, while InferCode \cite{buiInferCodeSelfSupervisedLearning2021} and TransformCode \cite{xian_2024} employ self-supervised learning to generate embeddings. In addition, we incorporated CodeBERT \cite{codebert2020} in an unsupervised setting, as well as LLM2Vec \cite{behnamghader2024llm2vec}, evaluated under two configurations: LLM2Vec [Llama-2-7B] (Bi + MNTP) and LLM2Vec [Llama-2-7B] (Bi + MNTP + Supervised).

Although both \textsc{LSem2Vec} and LLM2Vec employ large language models for code representation, they differ fundamentally in design. LLM2Vec fine‑tunes pre‑trained LLMs with contrastive or supervised objectives, requiring additional training resources and labeled datasets to adapt the model parameters for embedding tasks. In contrast, \textsc{LSem2Vec} avoids fine‑tuning altogether and instead leverages the inherent semantic capabilities of LLMs combined with sentence embedding techniques to generate source code embeddings in a purely unsupervised manner. This distinction highlights the originality of \textsc{LSem2Vec}, which emphasizes efficiency and generalizability without the computational overhead of fine‑tuning.

To measure the similarity between two code fragments, we used cosine similarity to compute the distance between their code embeddings, thereby keeping \textsc{LSem2Vec} unsupervised. 
Cosine similarity is selected because code-clone detection aims to measure the semantic similarity between two code representations. 
It focuses on the relative direction of embedding vectors and is commonly used in embedding-based retrieval and similarity comparison tasks.
It is important to emphasize that CodeBERT was not trained with a supervised clone-detection classifier, as incorporating one would have violated the unsupervised learning assumption integral to our experiment. Instead, CodeBERT, along with other models such as Code2vec, InferCode, and LLM2Vec, employs a similar prediction methodology to ours. In this methodology, the clone label is predicted based on the cosine similarity between the code embeddings of two fragments. By utilizing these methods, we ensure that \textsc{LSem2Vec} remains consistent with the principles of unsupervised learning while effectively identifying code clones.

In our ablation experiments on the two datasets, we compared \textsc{LSem2Vec} against the baseline LLM. In \textsc{LSem2Vec}, clone labels are predicted based on the cosine similarity between code embeddings. The baseline LLM generates a prompt utilizing the code pairs shown in Figure~\ref{old_prompt}. The LLM responds with ``Yes'' for clone detection or ``No'' for non-clone detection.

It is important to note that the baseline LLM queries are more time-consuming compared to \textsc{LSem2Vec}. This increased time is due to a single code snippet appearing multiple times across different code pairs, requiring the LLM to be called repeatedly for each occurrence. In contrast, \textsc{LSem2Vec} significantly optimizes the process by requiring only a single LLM call to generate embeddings, which can then be used for various downstream tasks. This efficiency is illustrated in Figure~\ref{fig_code_clone}.

For code clustering, we selected both text-based and code-specific baselines. Word2vec \cite{NIPS2013_word2vec} and Doc2vec \cite{doc2vec_2014} treat source code as text to generate embeddings, while the Sequential Denoising Auto Encoder (SAE) \cite{hill-etal-2016-learning} encodes and reconstructs text to capture structural semantics. Code-specific models such as Code2vec \cite{alon2019code2vec,kang2019assessing} and Code2seq \cite{alon2018code2seq} leverage abstract syntax trees to learn structured representations, whereas InferCode \cite{buiInferCodeSelfSupervisedLearning2021} applies self-supervised subtree prediction to obtain robust embeddings. LLM2Vec \cite{behnamghader2024llm2vec} was again included under the same two pre-trained settings as in clone detection. It is worth noting that baseline LLMs face limitations in clustering tasks due to context length constraints, as multiple code fragments may exceed the maximum allowable input size, leading to incomplete or inaccurate processing, as demonstrated in Figure~\ref{fig_out_of_limited}.

\subsection{Dependent Variables}
There are two dependent variables, relating to code-clone detection and code clustering tasks.
We followed the original studies \cite{buiInferCodeSelfSupervisedLearning2021,zhangNovelNeuralSource2019} to guide the choice of evaluation metrics for these experiments.

\subsubsection{Metrics for Code-Clone Detection}
Code clone detection is a classification task that determines whether two code fragments are identical.
To evaluate the performance of code-clone detection, we use the following metrics that are commonly used in classification tasks:
\begin{equation}
        Accuracy =  \left(\frac{TP+TN}{TP+TN+FP+FN}\right),
        \label{eq_acc}
\end{equation}
\begin{equation}
        Precision = \left(\frac{TP}{TP+FP}\right),
        \label{eq_prec}
\end{equation}
\begin{equation}
        Recall = \left(\frac{TP}{TP+FN}\right),
        \label{eq_recall}
\end{equation}
\begin{equation}
        F1\text{-}score = 2 \times \frac{Precision \times Recall}{Precision + Recall}.
        \label{eq_f1}
\end{equation}
where $TP$, $TN$, $FP$, and $FN$ represent the true positives, true negatives, false positives, and false negatives, respectively.
Accuracy measures the proportion of correct predictions among all predictions;
Precision measures the proportion of positive predictions that are positive;
Recall measures the proportion of positive instances that are correctly predicted; and
F1-score is the harmonic mean of precision and recall, which balances both metrics.

Accuracy may not be a reliable metric for imbalanced datasets \cite{imbalance_data_2013}, where some classes or categories are underrepresented or overrepresented. 
In such cases, Accuracy may be biased toward the dominant class, resulting in the minority class being ignored.
For example, if a dataset has 95\% positive instances and 5\% negative instances, a classifier that always predicts positive will achieve 95\% accuracy but will fail to detect any negative instances.
Therefore, Accuracy may not reflect the classifier's true performance on imbalanced datasets.
To address this, we use the F1-score (the harmonic mean of precision and recall) as an alternative metric for imbalanced datasets \cite{imbalance_data_2013}.
The F1-score takes into account both precision and recall:
It assigns a higher value when both precision and recall are high, indicating that the classifier can correctly identify both positive and negative instances.
The F1-score is lower when either precision or recall is low, indicating that the classifier either misses some positive instances or produces false positives.
In summary, the F1-score provides a more accurate and robust measure of the classifier's performance on imbalanced datasets.

\subsubsection{Metrics for Code Clustering}
We employ the adjusted Rand index (ARI) \cite{halkidi2001clustering} to evaluate model performance on code clustering tasks, a widely recognized and robust metric for assessing clustering quality \cite{xian_2021_AGGM,Xian2022}. 
Unlike the traditional Rand Index, ARI adjusts for the chance grouping of elements, providing a more accurate measure of clustering quality. 
This metric computes the similarity between two clusterings by considering all sample pairs and determining whether they are assigned to the same or different clusters in the predicted and true clusterings.
The ARI score ranges from -1 to 1, where 
1 indicates perfect agreement between the clusterings; 0 indicates random labeling; and negative values indicate less agreement than expected by chance. 
In our experiment, we employ the ARI to assess the effectiveness of various code clustering methods, ensuring robust validation of clustering performance.
The ARI is defined as:
\begin{equation}
\resizebox{\linewidth}{!}{$
\operatorname{ARI}\left(C_{truth}, C_{pred}\right)
=
\frac{
\sum_{ij}\binom{N_{ij}}{2}
-
\left[\sum_i\binom{N_i}{2}\sum_j\binom{N_j}{2}\right]/\binom{N}{2}
}{
\frac{1}{2}\left[\sum_i\binom{N_i}{2}+\sum_j\binom{N_j}{2}\right]
-
\left[\sum_i\binom{N_i}{2}\sum_j\binom{N_j}{2}\right]/\binom{N}{2}
}
$}
\label{eq_ari}
\end{equation}
where $N$ represents the total number of data points in the dataset. $C_{truth}$ denotes ground-truth clustering and $C_{pred}$ denotes predicted clustering.
$N_{ij}$ is the number of data points of the class label $C_j \in C_{truth}$ assigned to cluster $C_i$ in partition $C_{pred}$.
$N_i$ is the number of data points in cluster $C_i$ of partition $C_{pred}$, and $N_j$ is the number of data points in class $C_j$ of partition $C_{truth}$.

\subsection{Dataset Preprocessing and Embedding Configuration}
To ensure reproducibility and fair comparison, all datasets (POJ-104, OJClone C, BigCloneBench) were uniformly cleaned before being processed by LSem2Vec. Comments, redundant whitespace, and inconsistent indentation were removed, while syntactically broken or trivial fragments were discarded. For OJClone C, clone and non-clone pairs were sampled with a fixed random seed, and official dataset splits were strictly preserved. No syntax tree parsing or dependency graph construction was applied; only raw cleaned source text was used.

All large language models employed their native tokenizers, with strict adherence to context length limits. When inputs exceeded these limits, code was split by function boundaries and summarized separately, with only the first complete sentence of each LLM output retained as the code summary. For Chinese experiments, a programming-oriented stop-word list was applied to generate both raw and filtered summaries.

Sentence embeddings were generated using pre-trained models with fixed dimensionality: 384 for English and 768 for Chinese. All vectors were L2-normalized to stabilize similarity computations, and no dimensionality reduction was applied during inference. Summaries and embeddings were stored in JSON and NumPy formats to enable efficient reuse across tasks.
For all LLM-based experiments, the same prompt template was applied to different LLMs without additional demonstrations or task-specific instructions. 
No fine-tuning was performed for any LLM or sentence embedding model. 
The generated summaries were directly used as inputs for the corresponding sentence embedding models.

\subsection{Experimental Environment}
All experiments were conducted on a computer featuring an AMD Ryzen 7 5700X processor, dual Nvidia RTX 3090 GPUs, and 64GB of DDR4 RAM. 
The algorithm was implemented with Python 3.9.12. 
We utilized several Python libraries essential for our research, including PyTorch 11.3 (torch), Hugging Face's Transformers, and Sentence-Transformers. 

\begin{table*}[!t]
 \centering
 \caption{Metrics for \textsc{LSem2Vec} with different configurations using GPT-3.5 on the OJClone C dataset}
 \label{tab:ojclonec}
  \small
  \setlength\tabcolsep{1.2mm}
 \begin{tabular}{rcrrrrrrrrr}
 \hline
 \multirow{2}{*}{\textbf{Languages}} & & \multicolumn{3}{c}{\textbf{Architecture and Configuration}} & & \multicolumn{4}{c}{\textbf{Metrics}} \\ \cline{3-5} \cline{7-10}
 & & \textbf{LLMs} & \textbf{Embedding Models} & \textbf{Threshold (T)}&  & \textit{Accuracy} & \textit{Precision} & \textit{Recall} & \textit{F1-score} \\\hline
 English && GPT-3.5 & MiniLM-L12 & 0.75 & &81.81\% & 86.52\% & 81.81\% & 81.21\% \\
 English & &GPT-3.5 & MiniLM-L6 & 0.75 & &83.62\% & 87.37\% & 83.62\% & 83.19\% \\
 English && GPT-3.5 & MiniLM-L12 & 0.70 && 86.45\% & 89.04\% & 86.45\% & 86.22\% \\
 English & &GPT-3.5 & MiniLM-L6 & 0.70 && 88.01\% & 89.71\% & 88.01\% & 87.88\% \\
 English && GPT-3.5 & MiniLM-L12 & 0.65 && 89.67\% & 90.92\% & 89.67\% & 89.59\% \\
 English && GPT-3.5 & MiniLM-L6 & 0.65 && 90.36\% & 90.91\% & 90.36\% & 90.33\% \\
 English & &GPT-3.5 & MiniLM-L12 & 0.60 && 91.56\% & 91.92\% & 91.56\% & 91.54\% \\
 English && GPT-3.5 & MiniLM-L6 & 0.60 && 90.71\% & 90.72\% & 90.72\% & 90.72\% \\
 \rowcolor{lightblue} English & &GPT-3.5 & MiniLM-L12 & 0.55 && 91.83\% & 91.83\% & 91.83\% & 91.82\% \\
 English & &GPT-3.5 & MiniLM-L6 & 0.55 && 88.75\% & 89.12\% & 88.75\% & 88.73\% \\
 English && GPT-3.5 & MiniLM-L12 & 0.50 && {90.09\%} & {90.51\%} & {90.09\%} & {90.07\%} \\
 English & &GPT-3.5 & MiniLM-L6 & 0.50 & &84.70\% & 86.73\% & 84.70\% & 84.48\% \\ \hline\hline
 Chinese & &GPT-3.5 & SBERT-CN & 0.75 && 78.91\% & 81.25\% & 79.81\% & 78.50\% \\
 Chinese (no stop words) && GPT-3.5 & SBERT-CN & 0.75 && 77.35\% & 80.27\% & 77.35\% & 76.80\% \\
 \rowcolor{lightgrey} Chinese && GPT-3.5 & SBERT-CN & 0.70 && 79.40\% & 79.68\% & 79.40\% & 79.35\% \\
 Chinese (no stop words) & &GPT-3.5 & SBERT-CN & 0.70 && 78.19\% & 78.67\% & 78.19\% & 78.10\% \\
 Chinese && GPT-3.5 & SBERT-CN & 0.65 && 76.54\% & 76.80\% & 76.54\% & 76.48\% \\
 Chinese (no stop words) && GPT-3.5 & SBERT-CN & 0.65 && 76.59\% & 78.78\% & 76.59\% & 76.55\% \\
 Chinese && GPT-3.5 & SBERT-CN & 0.60 && 72.17\% & 74.48\% & 72.17\% & 71.49\% \\
 Chinese (no stop words) && GPT-3.5 & SBERT-CN & 0.60 && 72.73\% & 74.83\% & 72.73\% & 72.14\% \\
 Chinese & &GPT-3.5 & SBERT-CN & 0.55 && 66.96\% & 72.64\% & 66.96\% & 64.75\% \\
 Chinese (no stop words) && GPT-3.5 & SBERT-CN & 0.55 && 67.68\% & 73.32\% & 67.68\% & 65.61\% \\
 Chinese && GPT-3.5 & SBERT-CN & 0.50 && 61.69\% & 71.48\% & 61.69\% & 56.77\% \\
 Chinese (no stop words) && GPT-3.5 & SBERT-CN & 0.50 && 62.39\% & 72.02\% & 62.39\% & 57.78\% \\
 \hline
 \end{tabular}
\end{table*}

\section{Experimental Results and Discussions}
  \label{sec_experiment_result}

This section presents the experimental results of \textsc{LSem2Vec} in the code-clone detection and code clustering tasks, and answers to the research questions outlined in Section \ref{sec-rqs}.

\subsection{Results of Code-Clone Detection Tasks}
We initially present the results of code-clone detection tasks and answer the corresponding research questions.

\subsubsection{Answer to RQ1.1} 
We conducted the experiments on the original GPT-3.5 LLM (without training or fine-tuning) and the OJClone C dataset with both English and Chinese prompts.
Besides, we set a series of thresholds to compute the cosine similarity between code fragments.
Table \ref{tab:ojclonec} shows the evaluation results of \textsc{LSem2Vec} using different sentence-embedding models that support English and Chinese, respectively.
From the results, we have the following observations:
\begin{itemize}
    \item Even though the same LLM (GPT-3.5) was used, the choice of sentence embedding model and the threshold setting significantly affected the results.

    \item The MiniLM models for English consistently outperformed the SBERT-CN model for Chinese. 

    \item Within the MiniLM models, the 12-layer MiniLM model did not always yield the best results. When the threshold was set to 0.50, 0.55, or 0.60, the 6-layer MiniLM model performed better.
    
    \item For English, when the MiniLM-L12 sentence embedding model under the similarity threshold of 0.55 achieved the best performance, with an F1-score of 91.82\%. 
    
    \item For Chinese, the SBERT-CN sentence embedding model achieved an F1-score of 79.35\% when the similarity threshold was set to 0.7, indicating superior performance.

\end{itemize}

It is important to note that \textsc{LSem2Vec} is not limited to the sentence-embedding models used in these experiments. 
Moreover, we anticipate that fine-tuning the sentence embedding models could further enhance the performance. 

\begin{tcolorbox}[breakable,colframe=black,arc=1mm,left={1mm},top={1mm},bottom={1mm},right={1mm},boxrule={0.25mm}]
    \textit{\textbf{Summary of Answers to RQ1.1:}}
    Our observations indicate that, despite using the same LLM in \textsc{LSem2Vec}, the selection of the sentence embedding model significantly affects performance. 
    For example, the MiniLM models for English consistently outperformed the SBERT-CN model for Chinese. 
    As for the models that support the same language (i.e., MiniLM models), their performance also varies. 
\end{tcolorbox}

\subsubsection{Answer to RQ1.2}
To reveal the impact of removing stop words from Chinese code summaries on the performance of \textsc{LSem2Vec}.
We conducted an in-depth comparison between GPT-3.5 and the SBERT-CN sentence-embedding model on the OJClone C dataset. 
Table \ref{tab:ojclonec} presents the results of whether the stop words are removed from the Chinese code summaries.
Based on the results, we have the following observations:
\begin{itemize}
    \item Removing stop words from the LLM outputs did not significantly impact performance.

    \item When the threshold was set to 0.70 or 0.75, situations without stop words had slightly lower F1-scores. However, at thresholds 0.50, 0.55, 0.60, or 0.65, situations without stop words had slightly higher F1-scores, indicating a minor impact on the effectiveness of \textsc{LSem2Vec}.
    
\end{itemize}

\begin{tcolorbox}[breakable,colframe=black,arc=1mm,left={1mm},top={1mm},bottom={1mm},right={1mm},boxrule={0.25mm}]
    \textit{\textbf{Summary of Answers to RQ1.2:}}
    Stop words are typically considered non-essential in traditional NLP tasks. 
    However, based on experimental results, removing stop words sometimes led to a decline in performance. 
    This indicates that stop words still contain meaningful information about the context of the code summaries. 
    Therefore, careful consideration is necessary when deciding whether to remove stop words in code-related tasks to prevent potential adverse effects on performance.
\end{tcolorbox}

\begin{table*}[!t]
 \centering
 \caption{Metrics for \textsc{LSem2Vec} with different configurations using GLM on the OJClone C dataset}
 \label{tab:ojclonec_glm}
 \small
 \setlength\tabcolsep{1.2mm}
 \begin{tabular}{rcrrrrrrrrr}
 \hline
 \multirow{2}{*}{\textbf{Languages}} & & \multicolumn{3}{c}{\textbf{Architecture and Configuration}} & & \multicolumn{4}{c}{\textbf{Metrics}} \\ \cline{3-5} \cline{7-10}
 & & \textbf{LLMs} & \textbf{Embedding Models} & \textbf{Threshold (T)}&  & \textit{Accuracy} & \textit{Precision} & \textit{Recall} & \textit{F1-score} \\\hline
 \cline{2-8}
 \rowcolor{lightblue} English && GLM4 & MiniLM-L12 & 0.50 && 87.07\% & 88.03\% & 87.07\% & 86.99\% \\
 English && GLM4 & MiniLM-L6 & 0.50& & 86.60\% & 86.87\% & 86.60\% & 86.58\% \\
 Chinese && GLM4 & SBERT-CN & 0.70& & 72.02\% & 72.08\% & 72.02\% & 72.00\% \\
 Chinese (no stop words) && GLM4 & SBERT-CN & 0.70& & 71.75\% & 71.80\% & 71.75\% & 71.74\% \\
\hline
 \rowcolor{lightgrey} English && GLM3 & MiniLM-L12 & 0.50& & 78.62\% & 79.50\% & 78.62\% & 78.46\% \\
 English && GLM3 & MiniLM-L6 & 0.50& & 76.61\% & 76.79\% & 76.61\% & 76.57\% \\
 Chinese && GLM3 & SBERT-CN & 0.70& & 66.27\% & 66.87\% & 66.27\% & 65.97\% \\
 Chinese (no stop words) && GLM3 & SBERT-CN & 0.70& & 66.63\% & 67.90\% & 66.63\% & 66.02\% \\
 \hline
 \end{tabular}

\end{table*}

\subsubsection{Answer to RQ1.3}
Table \ref{tab:ojclonec} presents the results of \textsc{LSem2Vec} conducted by GPT-3.5.
To better evaluate the performance of \textsc{LSem2Vec} across different LLMs, we also conducted a series of experiments with the open-source GLM series LLMs on the OJClone C dataset.
Table \ref{tab:ojclonec_glm} shows the results of using GLM3 and GLM4 under the threshold of 0.50 and 0.70.
Based on the results, we have the following observations:
\begin{itemize}
    \item For English sentence embedding models, GLM4 achieved a higher F1-score of 86.99\% using the MiniLM-L12 embedding model with a threshold value of 0.5, indicating optimal performance. 
    
    \item For Chinese models, GLM4 still achieved the best performance at a threshold value of 0.7, with an F1-score of 72.00\%.
    The performance of \textsc{LSem2Vec} with GLM3 was worse than with GLM4, achieving only a 66.02\% F1-score.
    
    \item In general, \textsc{LSem2Vec} with GLM3 yielded slightly lower performance than GLM4.
    Among all cases, GLM3 achieved the highest F1-score of 78.46\% at a threshold of 0.5, and the MiniLM-L12 sentence embedding model did as well. 

    \item Compared to \textsc{LSem2Vec} with GPT-3.5, GLM4, and GLM3 required lower threshold values to attain their best performance.
    This discrepancy may be attributed to varying code summarization capabilities among LLMs, with GPT-3.5 demonstrating superior performance compared to GLM4 and GLM3 on this task.
\end{itemize}

\begin{tcolorbox}[breakable,colframe=black,arc=1mm,left={1mm},top={1mm},bottom={1mm},right={1mm},boxrule={0.25mm}]
    \textit{\textbf{Summary of Answers to RQ1.3:}}
    Our observations indicate that the selection of LLM plays a critical role in determining the overall performance of \textsc{LSem2Vec}.
    The effectiveness of \textsc{LSem2Vec} improves as more advanced and powerful LLMs develop.  
    Specifically, the GPT-3.5 consistently outperformed the other LLMs.
\end{tcolorbox}

\begin{table*}[!b]
 \centering
 \caption{Metrics for different methods on the OJClone C dataset (unsupervised)}
 \label{tab:ojclone_compare}
 \small
 \setlength{\tabcolsep}{5.1mm}
 \begin{tabular}{rcrrr}
 \hline
  \multirow{2}{*}{\textbf{Methods}} & \multicolumn{3}{c}{\textbf{Metrics}} \\  \cline{2-4}
 & \textit{Precision} & \textit{Recall} & \textit{F1-score} \\
\hline
 Deckard & \textbf{99.00\%} & 5.00\% & 10.00\% \\
 DLC & 71.00\% & 0.00\% & 0.00\% \\
 SourcererCC & 7.00\% & 74.00\% & 14.00\% \\
 Code2vec & 56.00\% & 69.00\% & 61.00\% \\
 CodeBERT & 77.48\% & 19.86\% & 16.43\% \\
 InferCode & 61.00\% & 70.00\% & 64.00\% \\
 TransformCode & 67.69\% & 67.29\% & 67.10\% \\
\hline
GPT-3.5 & 83.23\% & 78.68\% & 77.94\% \\
GLM4 & 79.98\% & 66.65\% & 62.47\% \\
GLM3 & 61.38\% & 59.89\% & 58.52\%  \\
\hline

 \rowcolor{lightgrey}LLM2Vec [Llama-2-7B] (Bi + MNTP) ($T=0.8$)  &  60.46\% & 51.45\%  & 38.13\% \\
LLM2Vec [Llama-2-7B] (Bi + MNTP) ($T=0.7$)  & 61.54\%  & 50.16\%  & 33.82\%  \\
LLM2Vec [Llama-2-7B] (Bi + MNTP) ($T=0.6$)  & 25.00\% & 50.00\% & 33.33\% \\

\rowcolor{lightgrey}LLM2Vec [Llama-2-7B] (Bi + MNTP + Supervised) ($T=0.8$) &  84.70\% & 83.80\%  &  83.70\% \\
LLM2Vec [Llama-2-7B] (Bi + MNTP + Supervised) ($T=0.7$) &  77.56\% & 68.76\%  &  66.05\% \\
LLM2Vec [Llama-2-7B] (Bi + MNTP + Supervised) ($T=0.6$) &  74.85\% & 52.24\%  &  38.19\% \\

\hline
 \rowcolor{lightblue} \textsc{LSem2Vec} (GPT-5.4 \& MiniLM-L12) & 95.59\% & \textbf{95.37\%} & \textbf{95.36\%} \\
 \rowcolor{lightblue} \textsc{LSem2Vec} (GPT-3.5 \& MiniLM-L12) & 91.83\% & {91.83\%} & {91.82\%} \\
 \rowcolor{lightgrey} \textsc{LSem2Vec} (GLM4 \& MiniLM-L12) & 88.03\% & 87.07\% & 86.99\% \\
 \rowcolor{lightgrey} \textsc{LSem2Vec} (GLM3 \& MiniLM-L12) & 79.50\% & 78.62\% & 78.46\% \\
 \hline
 \end{tabular}
 
\end{table*}

\begin{table*}[!t]
 \centering
 \caption{Metrics for different methods on the BigCloneBench dataset (unsupervised)}
 \label{tab:clone_compare}
  \small
\setlength{\tabcolsep}{5.1mm}
 \begin{tabular}{rcrrr}
 \hline
  \multirow{2}{*}{\textbf{Methods}} & \multicolumn{3}{c}{\textbf{Metrics}} \\  \cline{2-4}
 & \textit{Precision} & \textit{Recall} & \textit{F1-score} \\
\hline
 Deckard & 93.00\% & 2.00\% & 3.00\% \\
 DLC & \textbf{95.00\%} & 1.00\% & 1.00\% \\
 SourcererCC & 88.00\% & 2.00\% & 3.00\% \\
 Code2vec & 82.00\% & 40.00\% & 60.00\% \\
 CodeBERT & 77.48\% & 19.86\% & 16.43\% \\
 InferCode & 90.00\% & 56.00\% & 75.00\% \\
 TransformCode & 84.76\% & 87.50\% & 82.36\% \\ 
 \hline
GPT-3.5  & 81.75\% & 86.38\% & 81.42\% \\
GLM4 & 79.85\% & 85.48\% & 81.99\% \\
GLM3 & 82.25\% & 83.78\% & 82.87\%  \\
 \hline
\rowcolor{lightgrey}LLM2Vec [Llama-2-7B] (Bi + MNTP) ($T=0.8$) &  82.85\% & 19.74\%  &  15.60\% \\
LLM2Vec [Llama-2-7B] (Bi + MNTP) ($T=0.7$) &  83.48\% & 14.04\%  &  4.26\% \\
LLM2Vec [Llama-2-7B] (Bi + MNTP) ($T=0.6$) &  1.84\% & 13.56\%  &  3.23\% \\

LLM2Vec [Llama-2-7B] (Bi + MNTP + Supervised) ($T=0.8$) &  83.05\% & 87.97\%  &  83.63\% \\
\rowcolor{lightgrey} LLM2Vec [Llama-2-7B] (Bi + MNTP + Supervised) ($T=0.7$) &  86.01\% & 87.02\%  &  85.80\% \\
LLM2Vec [Llama-2-7B] (Bi + MNTP + Supervised) ($T=0.6$) &  83.79\% & 73.03\%  &  76.71\% \\

\hline
 \rowcolor{lightblue}\textsc{LSem2Vec} (GPT-5.4 \& MiniLM-L6) & 88.86\% & \textbf{89.79}\% & \textbf{86.50\%} \\
 
 \rowcolor{lightblue}\textsc{LSem2Vec} (GPT-3.5 \& MiniLM-L6) & 86.23\% & 87.27\% & {85.83\%} \\
 \rowcolor{lightgrey}\textsc{LSem2Vec} (GLM4 \& MiniLM-L6) & 85.35\% & 88.03\% & 85.19\% \\
 \rowcolor{lightgrey}\textsc{LSem2Vec} (GLM3 \& MiniLM-L6) & 84.41\% & 86.39\% & 85.06\% \\
 \hline
 \end{tabular}
\end{table*}

\subsubsection{Answer to RQ1.4}
Table \ref{tab:ojclone_compare} and Table \ref{tab:clone_compare} present the results of \textsc{LSem2Vec} and its baseline LLMs on two programming language datasets: OJClone for C and BigCloneBench for Java.
Based on the experimental results, we can observe the following:
\begin{itemize}
    \item In the GLM3 configuration, \textsc{LSem2Vec} surpasses its baseline, GLM3, on both the OJClone C and BigCloneBench datasets, achieving F1-scores of 78.46\% and 85.06\%, respectively.

    \item In the GLM4 configuration, \textsc{LSem2Vec} outperforms its baseline, GLM3, on both the OJClone C and BigCloneBench datasets, achieving F1-scores of 86.99\% and 85.19\%, respectively.

    \item In the GPT-3.5 configuration, \textsc{LSem2Vec} outperforms its baseline, GLM3, on both the OJClone C and BigCloneBench datasets, achieving F1-scores of 91.82\% and 85.83\%, respectively.

    \item In the GPT-5.4 configuration, \textsc{LSem2Vec} outperforms GPT-3.5, on both the OJClone C and BigCloneBench datasets, achieving F1-scores of 95.36\% and 86.50\%, respectively.
\end{itemize}

\begin{tcolorbox}[breakable,colframe=black,arc=1mm,left={1mm},top={1mm},bottom={1mm},right={1mm},boxrule={0.25mm}]
    \textit{\textbf{Summary of Answers to RQ1.4:}}
\textsc{LSem2Vec} surpasses its baseline LLMs in code-clone detection across both datasets, demonstrating particularly strong performance on the OJClone C dataset. Despite using refined prompts specifically designed for code-clone detection, the baseline LLMs still achieve lower performance than \textsc{LSem2Vec}.
\end{tcolorbox}

\subsubsection{Answer to RQ2}
Table \ref{tab:ojclone_compare} and Table \ref{tab:clone_compare} present the results of \textsc{LSem2Vec} on two programming language datasets: OJClone for C and BigCloneBench for Java.
Based on the results, we have the following observations:
\begin{itemize}
    \item For the OJClone C dataset (for C), \textsc{LSem2Vec} achieved the F1-scores range from 78.46\% to 91.82\%:
    78.46\% for the GLM3-based approach;
    86.99\% for the GLM4-based approach;
    91.82\% for the GPT-3.5-Turbo-based approach; and
    95.36\% for the GPT-5.4-based approach.

    \item For the BigCloneBench dataset (for Java), \textsc{LSem2Vec} achieved the F1-scores range from 85.06\% to 85.83\%:
    85.06\% for the GLM3-based approach;
    85.19\% for the GLM4-based approach; 
    85.83\% for the GPT-3.5-Turbo-based approach; and
    86.50\% for the GPT-5.4-based approach.

    \item The performance of \textsc{LSem2Vec} may vary by programming language: it achieved higher F1-scores on the C-based dataset than on the Java-based dataset.
    This discrepancy may be attributed to the inherent complexity of Java code compared to C code. 
    Java code with more complex structures presents additional challenges for LLMs in accurately summarizing and detecting code clones.
\end{itemize}

\begin{tcolorbox}[breakable,colframe=black,arc=1mm,left={1mm},top={1mm},bottom={1mm},right={1mm},boxrule={0.25mm}]
    \textit{\textbf{Summary of Answers to RQ2:}}
    \textsc{LSem2Vec} enhances the efficiency of code embedding generation and ensures embeddings are functional and relevant across programming languages.
    This is accomplished by leveraging LLMs' zero-shot learning capabilities to generate accurate, concise summaries of these code fragments. 
    Unlike traditional methods that may require extensive training data for each language, \textsc{LSem2Vec} leverages LLMs' inherent ability to produce broad, adaptable code summaries. 
\end{tcolorbox}

\subsubsection{Answer to RQ3: Code-Clone Detection Perspective}
\label{sec-rq3-result-clone}
Table \ref{tab:ojclone_compare} and Table \ref{tab:clone_compare} also compare the performance between \textsc{LSem2Vec} and other methods.
Based on the results, we have the following observations:
\begin{itemize}
    \item Regardless of the programming languages, \textsc{LSem2Vec} always achieved the best F1-scores, significantly surpassing the state-of-the-art unsupervised method, TransformCode.

    \item \textsc{LSem2Vec} using GPT-3.5 and GPT-5.4 achieved higher F1-scores than the GLM series LLMs.

    \item In comparison to the unsupervised LLM2Vec trained on public data, \textsc{LSem2Vec} consistently surpassed it across various thresholds on two datasets. The supervised version of LLM2Vec, on the other hand, requires substantial computational resources, and training on 8 NVIDIA A100 GPUs with an effective batch size of 512 for 1000 steps \cite{behnamghader2024llm2vec}. Despite this high resource consumption, \textsc{LSem2Vec} still outperforms the supervised LLM2Vec, particularly on the OJClone C dataset.
\end{itemize}

To evaluate the performance of \textsc{LSem2Vec} against state-of-the-art methods, we also conducted a series of experiments using different approaches from a code-clustering perspective. 
The detailed results are presented and analyzed in Section \ref{sec-rq3-result-clustering}. 
After analyzing these two perspectives, we will answer \textbf{RQ3}.

\begin{table*}[!t]
 \centering
 \caption{Result of code clustering in adjusted rand index (ARI) on OJClone C dataset (unSupervised)}
 \label{tab:code_cluster}
 \setlength{\tabcolsep}{10mm}
 \small
 \begin{tabular}{rcr}
 \hline
\multirow{2}{*}{\textbf{Methods}} & \multicolumn{1}{c}{\textbf{Metric}} \\
 \cline{2-2}
 &  \textit{ARI} \\
 \hline
 Word2vec &  0.28 \\
 Doc2vec &  0.42 \\
 SAE & 0.41 \\
 Code2vec  & 0.58 \\
 Code2seq  & 0.53 \\
 InferCode & 0.70 \\\hline
LLM2Vec [Llama-2-7B] (Bi + MNTP) & 0.30 \\  
LLM2Vec [Llama-2-7B] (Bi + MNTP + Supervised)  & 0.90 \\  
 \hline
 \rowcolor{lightgrey}\textsc{LSem2Vec} (GLM3 \& MiniLM-L6) &  0.51 \\
 \rowcolor{lightgrey}\textsc{LSem2Vec} (GLM3 \& MiniLM-L12) &  0.47 \\
 \rowcolor{lightgrey}\textsc{LSem2Vec} (GLM4 \& MiniLM-L6) &  0.78 \\
 \rowcolor{lightgrey}\textsc{LSem2Vec} (GLM4 \& MiniLM-L12) &  0.89 \\
 \rowcolor{lightgrey}\textsc{LSem2Vec} (GPT-3.5 \& MiniLM-L6) &  {0.97} \\
 \rowcolor{lightgrey}\textsc{LSem2Vec} (GPT-3.5 \& MiniLM-L12) &  0.91 \\

  \rowcolor{lightblue}\textsc{LSem2Vec} (GPT-5.4 \& MiniLM-L6) &  0.98 \\
  \rowcolor{lightblue}\textsc{LSem2Vec} (GPT-5.4 \& MiniLM-L12) &  \textbf{0.99} \\
 \hline
 \end{tabular}
\end{table*}

\subsection{Results of Code Clustering Tasks}
This section presents the results of code-clone detection tasks and answers the corresponding research questions.

\subsubsection{Answer to RQ3: Code Clustering Perspective}
\label{sec-rq3-result-clustering}
Table \ref{tab:code_cluster} shows the ARI results of \textsc{LSem2Vec} and other approaches in code clustering tasks.
Table~\ref{tab:code_cluster} also presents a comparative analysis of \textsc{LSem2Vec} against various configurations of LLMs and embedding models.
Based on the results, we have the following observations:
\begin{itemize}
    \item 
    \textsc{LSem2Vec} with GPT-5.4 demonstrated superior performance using the MiniLM-L12 sentence embedding model. 
    Despite using the same sentence-embedding models, \textsc{LSem2Vec} performs significantly better with more advanced LLMs. 

    \item 
    \textsc{LSem2Vec} with GLM4 surpassed the state-of-the-art approach InferCode, achieving an ARI of 0.89.

    \item 
    \textsc{LSem2Vec}, utilizing GPT-5.4, achieved an impressive ARI of 0.99. This performance surpasses the supervised LLM2Vec, which achieved an ARI of 0.90, making \textsc{LSem2Vec} the most effective among all evaluated methods.
\end{itemize}

\begin{tcolorbox}[breakable,colframe=black,arc=1mm,left={1mm},top={1mm},bottom={1mm},right={1mm},boxrule={0.25mm}]
    \textit{\textbf{Summary of Answers to RQ3:}}
    We conducted a comparative analysis of \textsc{LSem2Vec} against other unsupervised methods on two tasks: code-clone detection (Section \ref{sec-rq3-result-clone}) and code clustering (Section \ref{sec-rq3-result-clustering}). 
    Our observations indicate that \textsc{LSem2Vec} demonstrates a significant advantage, particularly in the code-clone detection and code clustering tasks. 
    Unlike other approaches, \textsc{LSem2Vec} requires no training phase or labeled data. 
    Instead, it directly generates code embeddings from code fragments, contributing to high efficiency. 
    This efficiency is evident compared to other unsupervised methods, making \textsc{LSem2Vec} a superior choice for these tasks.
\end{tcolorbox}

\begin{figure*}[!t]
        \centering
        \graphicspath{{materials_img/}}

        \subfigure[\textsc{LSem2Vec} (GPT-5.4 + MiniLM-L12)]{
        \label{fig_tsne_a}
                \includegraphics[width=0.315\textwidth]{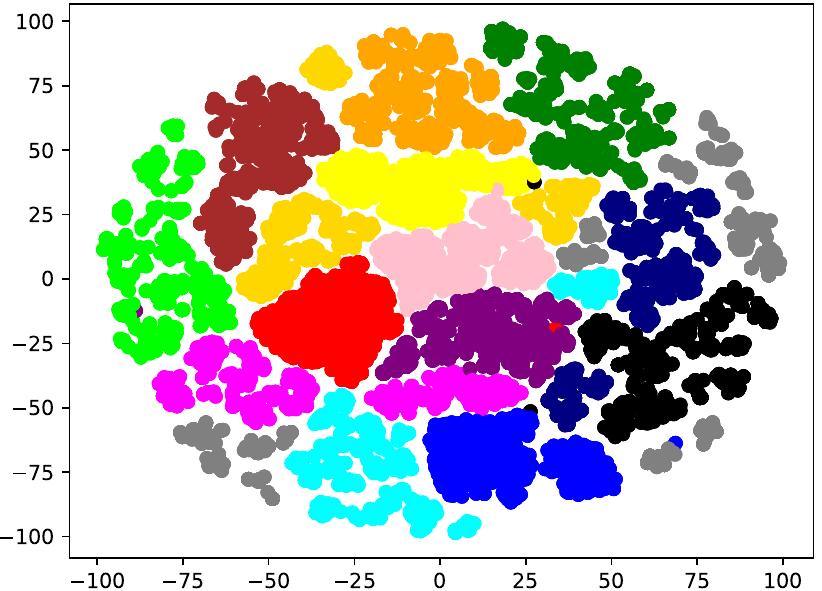}
        }
        \subfigure[\textsc{LSem2Vec} (GPT-5.4 + MiniLM-L6)]{
        \label{fig_tsne_b}
                \includegraphics[width=0.315\textwidth]{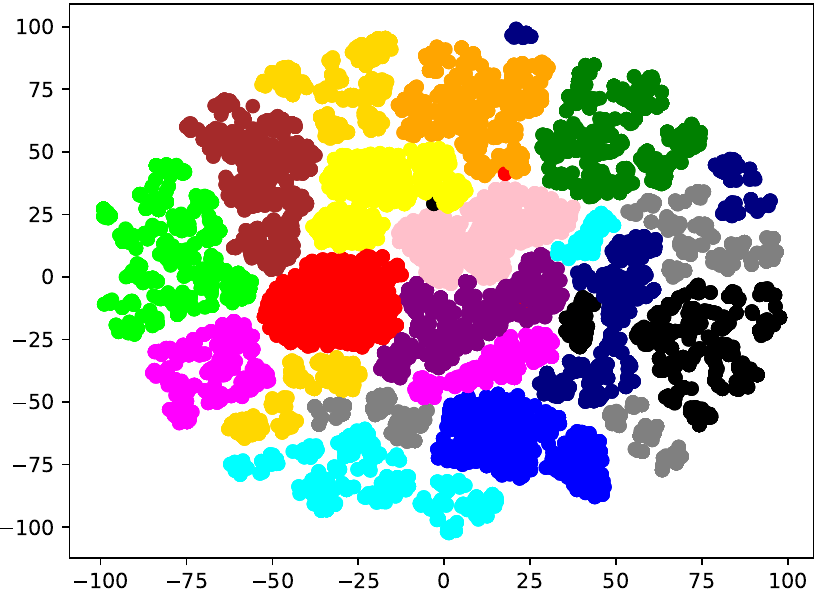}
        }
        \subfigure[\textsc{LSem2Vec} (GPT-3.5 + MiniLM-L12)]{
        \label{fig_tsne_a}
                \includegraphics[width=0.315\textwidth]{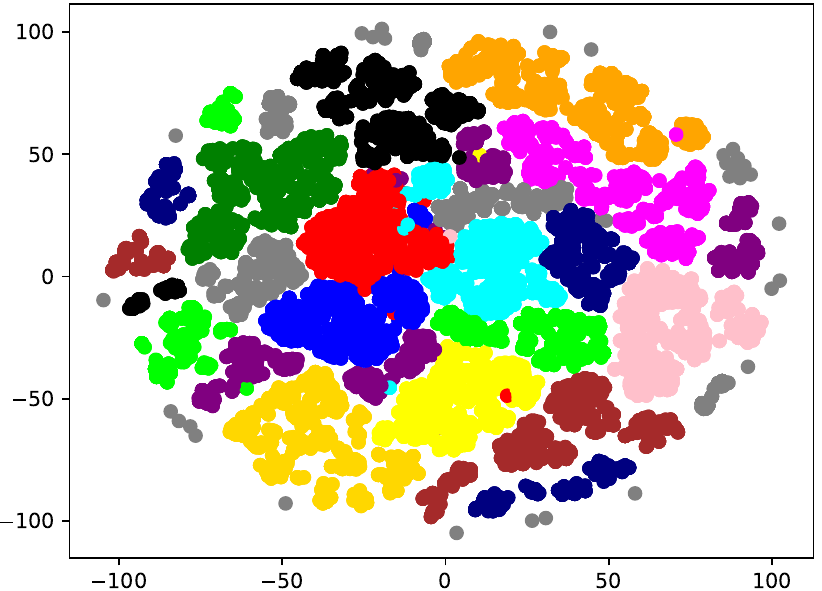}
        }
        
        \subfigure[\textsc{LSem2Vec} (GPT-3.5 + MiniLM-L6)]{
        \label{fig_tsne_b}
                \includegraphics[width=0.315\textwidth]{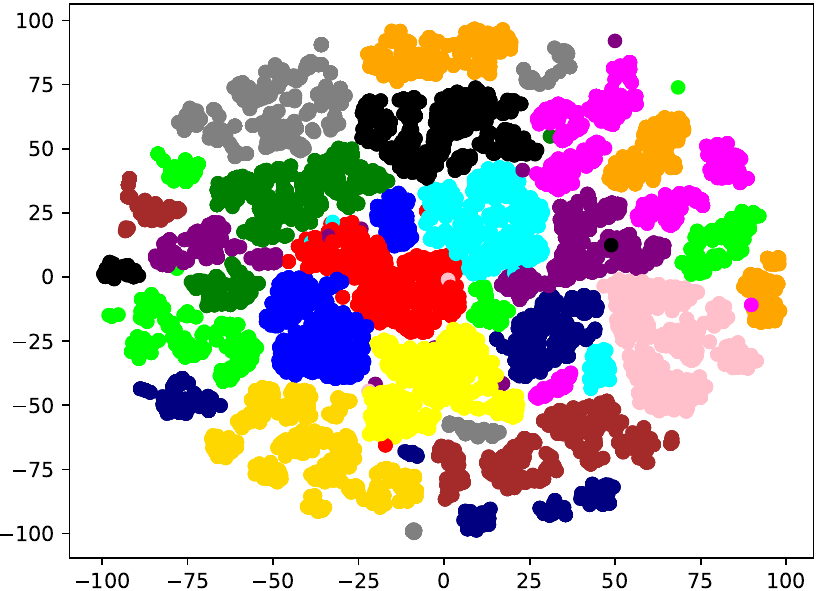}
        }
        \subfigure[\textsc{LSem2Vec} (GLM4 + MiniLM-L12)]{
        \label{fig_tsne_c}
                \includegraphics[width=0.315\textwidth]{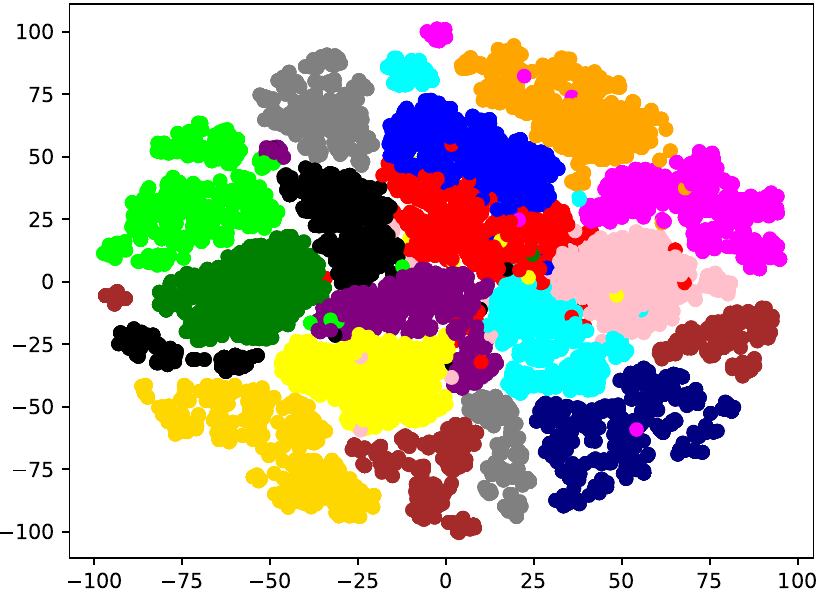}
        }
        \subfigure[\textsc{LSem2Vec} (GLM4 + MiniLM-L6)]{
        \label{fig_tsne_d}
                \includegraphics[width=0.315\textwidth]{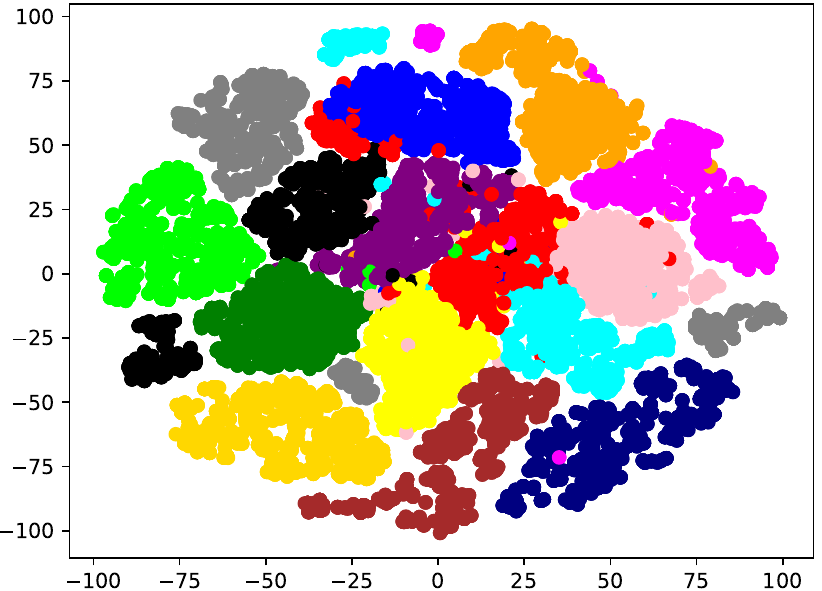}
        }     
        
        \subfigure[\textsc{LSem2Vec} (GLM3 + MiniLM-L12)]{
        \label{fig_tsne_e}
                \includegraphics[width=0.315\textwidth]{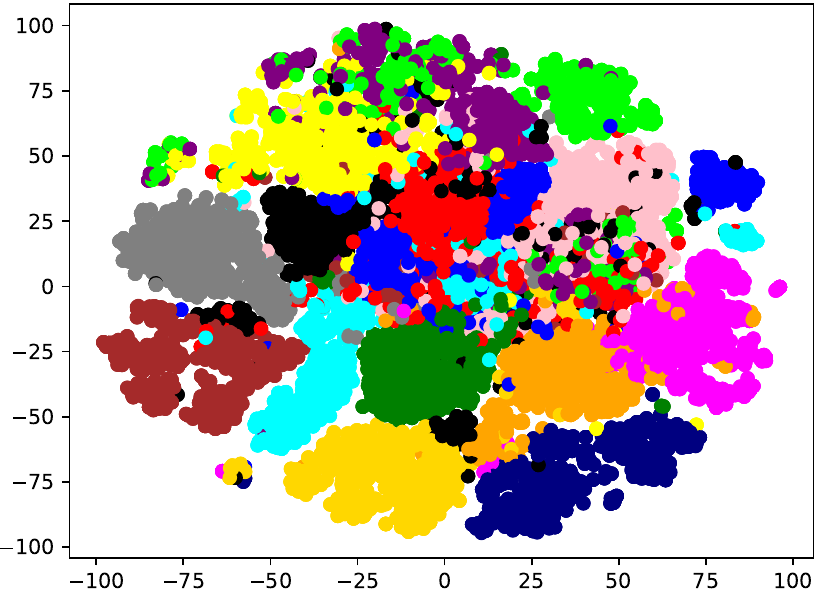}
        }
        \hspace{0.03\textwidth}
        \subfigure[\textsc{LSem2Vec} (GLM3 + MiniLM-L6)]{
        \label{fig_tsne_f}
                \includegraphics[width=0.315\textwidth]{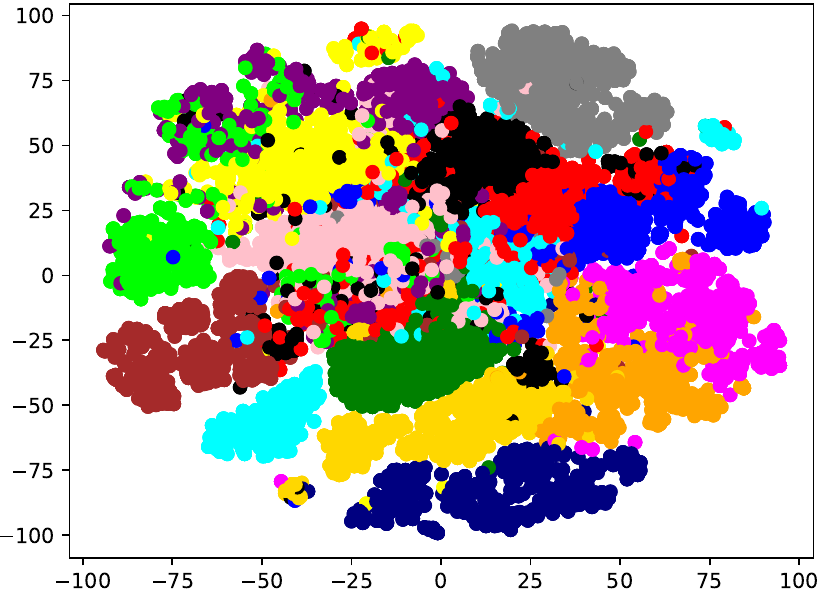}
        }
        
        \subfigure[LLM2Vec (Bi + MNTP)]{
        \label{fig_tsne_g}
                \includegraphics[width=0.315\textwidth]{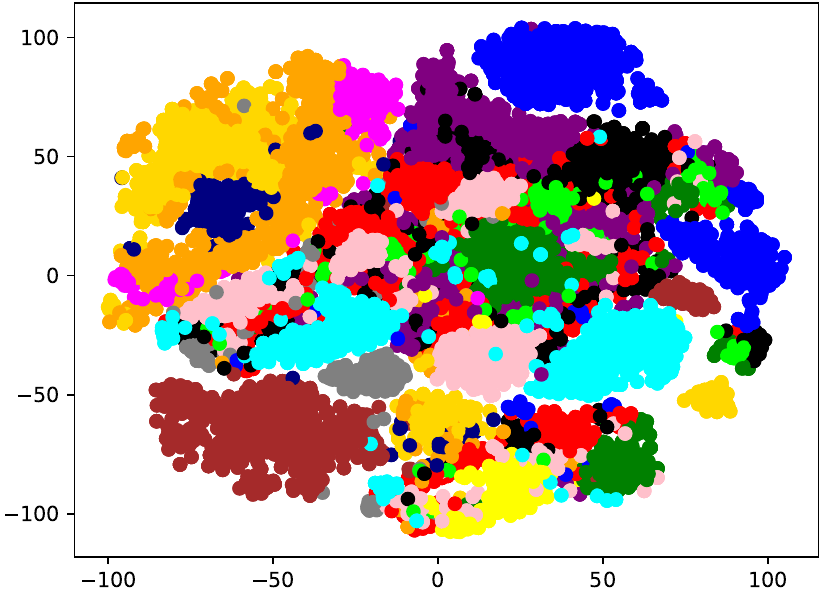}
        }
        \hspace{0.03\textwidth}
        \subfigure[LLM2Vec (Bi + MNTP + Supervised)]{
        \label{fig_tsne_h}
                \includegraphics[width=0.315\textwidth]{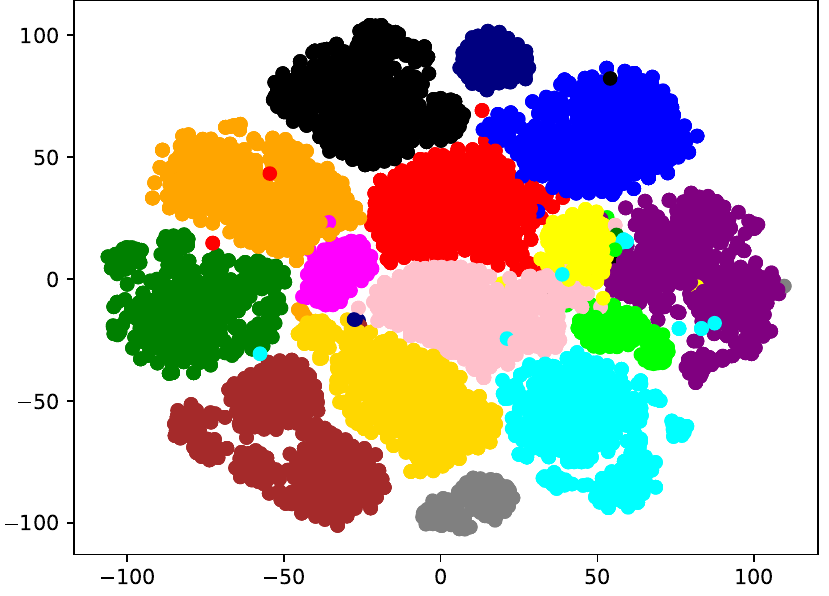}
        }
        \caption{Visualization with {t-SNE} using different configurations of \textsc{LSem2Vec} and LLM2Vec}
        \label{fig_tsne}
\end{figure*}

\subsubsection{Answer to RQ4}
To better understand why \textsc{LSem2Vec} achieved high ARI scores, we visualized the code embeddings it generated. 
Specifically, these embeddings are complex representations that capture both the structure and meaning of code fragments. 
We used t-distributed stochastic neighbor embedding (t-SNE), a widely used technique for dimensionality reduction and visualization of high-dimensional data, to visualize the complex embeddings.
This helps us to capture the similarities and differences of the generated code embeddings more easily.
We used t-SNE to project the multi-dimensional code vectors into a two-dimensional space and subsequently plotted the projections using Matplotlib \cite{matplotlib_Hunter_2007}. 

As a non-linear dimensionality reduction technique, t-SNE effectively preserves the local structure and relative distances of data points in the reduced-dimensional space, enabling accurate interpretation of relationships between code fragments in the visualized space. 
Specifically, t-SNE can capture complex patterns and relationships among code embeddings, convert similarities into joint probabilities, and then minimize the Kullback-Leibler divergence between these joint probabilities in the high- and low-dimensional spaces.
This process results in a two-dimensional representation that maintains the meaningful structure of the original data: 
Similar code fragments close to each other in high-dimensional space remain close in two-dimensional space, while different code fragments are farther apart in two-dimensional space.

Figure \ref{fig_tsne} provides clear t-SNE visualizations of the code clustering capabilities of \textsc{LSem2Vec} with various LLM and sentence embedding model configurations.
Based on the results (Figure \ref{fig_tsne} and Table~\ref{tab:code_cluster}), we have the following observations:
\begin{itemize}
    \item \textsc{LSem2Vec} effectively groups similar code fragments together with different LLMs.

    \item \textsc{LSem2Vec} with GLM3 resulted in ARIs of 0.51 and 0.47, indicating poorly defined cluster boundaries and significant overlap among clusters.

    \item \textsc{LSem2Vec} using GPT-3.5 achieved higher ARIs of 0.97 and 0.91, indicating well-defined, distinct cluster boundaries.

    \item \textsc{LSem2Vec} using GPT-5.4 achieved the best ARIs of 0.99 and 0.98, indicating the best cluster boundaries.

    \item \textsc{LSem2Vec} with GLM4 demonstrated performance comparable to \textsc{LSem2Vec} with GPT-3.5, achieving similarly high ARIs and clear cluster separations.

    \item 
    The supervised LLM2Vec significantly outperforms the unsupervised LLM2Vec, indicating a strong reliance on supervised training to generate better code embeddings.

    \item Better performance and higher ARIs are associated with more explicit boundaries in the t-SNE plots. 
    For example, Figure~\ref{fig_tsne_a} exhibits much clearer boundaries than Figure~\ref{fig_tsne_c}.
    Meanwhile, the ARI of Figure~\ref{fig_tsne_a} was 0.97, whereas Figure~\ref{fig_tsne_c} had an ARI of only 0.51. 
    
\end{itemize}

\begin{tcolorbox}[breakable,colframe=black,arc=1mm,left={1mm},top={1mm},bottom={1mm},right={1mm},boxrule={0.25mm}]
    \textit{\textbf{Summary of Answers to RQ4:}}
    The visualizations reveal that higher-quality code embeddings achieve less overlap and more distinct boundaries, consistent with the ARI metric findings (Section \ref{sec-rq3-result-clustering}).
    At the same time, the selection of LLM significantly impacts the effectiveness and quality of the code embeddings generated by \textsc{LSem2Vec}. 
    When properly configured, \textsc{LSem2Vec} can achieve high-quality code embeddings with well-defined clusters: A more advanced LLM within \textsc{LSem2Vec} leads to significantly improved code embeddings. 
\end{tcolorbox}

\section{Threats to Validity}
\label{sec:threatstovalidity}

This section explores key factors that may threaten the validity of the experimental results for our approach.

The first threat is related to the LLMs we selected.
We have employed four LLMs (i.e., GPT-3.5 Turbo, GPT-5.4, GLM3, and GLM4) in our proposed approach.
Meanwhile, our approach is flexible for the extension of integrating more proprietary LLMs, when working with large datasets. 
The cost associated with these models can be substantial due to the high price of tokens. 
Nevertheless, our approach is more cost-effective compared to other methods that utilize LLMs directly for downstream tasks. 
This is because our method requires generating the code summary only once (as illustrated in Section \ref{sec-code-summaries and storage}), allowing for multiple uses of the summary across various downstream tasks on a specific dataset.
Another threat concerns the stochastic nature of LLM-generated summaries. Different runs may produce slightly different wording even for the same code fragment, which could affect the resulting sentence embeddings. To reduce this threat, we used a fixed prompt template and fixed decoding configuration for each model throughout the experiments, retained only the first complete sentence as the summary, and stored generated summaries before downstream evaluation so that clone detection and clustering were run on a stable summary set.
Another threat is related to the quality of LLM-generated summaries.
Since these summaries serve as intermediate representations in \textsc{LSem2Vec}, inaccurate or incomplete summaries may introduce noise into the resulting embeddings.
Although our approach reduces randomness through a fixed generation process, the effectiveness of \textsc{LSem2Vec} still depends on the semantic understanding capability of the selected LLMs.

The second issue concerns the use of pre-trained sentence embedding models. Although \textsc{LSem2Vec} uses these models to convert sentences into embeddings, the models themselves are not trained or fine-tuned on datasets related to code. This means they might not perform as well as they could. Using embedding models that are specifically trained on code-related datasets could potentially improve the performance of our framework. These specialized models would likely capture the nuances and specific characteristics of programming languages better, leading to more accurate and effective embeddings for our purposes.

The third challenge relates to the effectiveness and cost of our proposed method. Our experiments show that our method performs better when higher-quality LLMs or sentence-embedding models are used. However, using high-end LLMs can be very expensive due to the cost per token. Additionally, these advanced LLMs cannot be deployed locally and must be accessed through an API, which requires a stable network connection. This reliance on an external network can be a limitation compared to open-source LLMs that can be installed and run on local machines. Using open-source LLMs is generally more cost-effective and does not depend on a network, making them a practical alternative in network-limited environments.

Another important consideration is that source code often exceeds the context length of current LLMs, which poses a threat to the validity of embedding-based approaches. To address this limitation, we plan to explore several mitigation strategies. One approach is to summarize code segments into shorter representations before embedding, thereby preserving essential semantics while reducing input size. Additionally, hierarchical encoding methods can be employed, where functions or modules are first summarized individually and then combined into higher-level embeddings. We also envision integrating retrieval-augmented generation, enabling the model to dynamically access relevant code fragments beyond its immediate context window. Finally, incorporating program-level structures such as function call graphs can provide complementary information that reduces reliance on long raw sequences. Together, these strategies strengthen the robustness of \textsc{LSem2Vec} and ensure its applicability even when handling large-scale software systems.

The last important issue is that LLMs sometimes produce errors or do not respond due to the issue of out-of-context windows, even when we use late interaction methods (see Section~\ref{sec_late_interaction}). In such cases, if the LLM fails to generate any output, our system cannot obtain the source code embeddings. To handle this, we plan to retry the request several times. If the LLM still fails, we can use backup models with larger context windows to increase the chance of receiving a valid response. This fallback strategy should help us obtain the source code embeddings as expected. However, it remains difficult to automatically detect and identify the errors in the LLM's response, which is an area that requires further research.

\section{Conclusions and Future Work}
\label{sec_conclusion}
In this paper, we propose a simple yet effective two-stage approach, \textsc{LSem2Vec} (\textit{\underline{\textbf{L}}LM-extracted code \underline{\textbf{Sem}}antics \underline{\textbf{to}} \underline{\textbf{Vec}}tor embedding}), to source code embedding generation. 
To the best of our knowledge, this is the first study to apply large language models and sentence embedding models to generate source code embeddings. 
\textsc{LSem2Vec} leverages the zero-shot learning capabilities of a large language model (LLM) to effectively summarize source code fragments, and then generate meaningful source code embeddings using a sentence embedding model.
We comprehensively evaluated \textsc{LSem2Vec} across several datasets and two tasks: source code clone detection and source code clustering. 
The experimental results directly demonstrate the superiority of \textsc{LSem2Vec} over existing approaches. 
Specifically, \textsc{LSem2Vec} outperformed 16 established methods for unsupervised source code clone detection, including notable ones such as SourcererCC \cite{sajnani2016sourcerercc}, Code2vec \cite{alon2019code2vec,kang2019assessing}, InferCode \cite{buiInferCodeSelfSupervisedLearning2021}, TransformCode \cite{xian_2024}, and LLM2Vec \cite{behnamghader2024llm2vec}.
The advantages of \textsc{LSem2Vec} include various aspects.
First, it requires no training or fine-tuning, significantly reducing the computational resources and time required for model preparation. 
Second, \textsc{LSem2Vec} is designed to be in various steps, allowing it to be decoupled into smaller components. 
These smaller components enable seamless application across various datasets and adaptation to different LLMs and sentence-embedding model configurations. 
Additionally, \textsc{LSem2Vec} includes a storage mechanism for dataset-specific source code summaries, which can be used for downstream tasks. 
The inherent flexibility of \textsc{LSem2Vec} ensures easy integration into diverse source-code analysis workflows, providing robust and efficient source-code embeddings without the need for domain-specific training or fine-tuning.

\textsc{LSem2Vec} offers a new direction for using an LLM to generate source code embeddings, with significant implications for future research and practical applications in this field. By leveraging an LLM, we provide an effective way to capture the meaning of source code without requiring extensive manual feature engineering or domain-specific training.
In our future work, we plan to expand the use of this method to solve more complex and practical SE challenges by incorporating a decoder component. This addition will make it more adaptable to a wider range of tasks. Specifically, we aim to create embeddings that are not only robust but also generalizable across various SE applications. These include code summarization \cite{gaoCodeStructureGuided2022,tangASTtransCodeSummarization2022,liuRetrievalAugmentedGenerationCode2022}, which helps in improving code readability and comprehension, and vulnerability detection \cite{wuVulCNNImageinspiredScalable2022,chakrabortyDeepLearningBased2020}, which plays a crucial role in identifying security risks in software. 
Except that, we also plan to investigate how LLM-generated embeddings can be applied to binary code analysis \cite{lu2025progressive,li2024rcfg2vec,jiang2024bincola}. Unlike source code, binary code lacks explicit syntax and structure, making traditional analysis more challenging. To enhance the effectiveness of \textsc{LSem2Vec}, we plan to incorporate additional information into the model, including function call graphs. These graphs capture relationships among functions within a program, helping the LLM better understand how different parts of the code are interconnected.

Moreover, we envision several concrete extensions to broaden the impact of \textsc{LSem2Vec}. First, developing multilingual embeddings will allow cross-language software engineering tasks, supporting diverse programming ecosystems. Second, integrating with retrieval-augmented generation can improve downstream tasks such as automated documentation and bug fixing by combining embeddings with external knowledge sources. Finally, deploying \textsc{LSem2Vec} within industrial SE pipelines will demonstrate its scalability and practical utility, bridging academic research with real-world software development practices. These directions highlight the broader potential of our approach to advance both theoretical understanding and applied solutions in software engineering.

 \section*{Acknowledgments}
 This work in this paper was partly supported by the Science and Technology Development Fund of Macau, Macao SAR, under Grant Nos. 0069/2025/RIB2 and 0021/2023/RIA1.

 \section*{Competing Interests}
 The authors declare that they have no competing interests or financial conflicts to disclose.

~\\

\balance
\bibliographystyle{IEEEtran}
\bibliography{references}

\clearpage
\newpage


\vfill

\end{document}